\documentclass[a4paper,11pt]{article} 
\pdfoutput=1

\usepackage{jcappub}
\usepackage{ulem}

\title{On ultra-high energy cosmic ray acceleration at the termination
  shock of young pulsar winds}

\author[a]{Martin Lemoine}
\author[a]{Kumiko Kotera}
\author[b]{J\'er\^ome P\'etri}

\affiliation[a]{Institut d'Astrophysique de Paris, \\CNRS,
  Universit\'e Pierre \& Marie Curie,\\ 98 bis boulevard Arago,
  F-75014 Paris, France} 
\affiliation[b]{Observatoire Astronomique de
  Strasbourg,\\ Universit\'e de Strasbourg, CNRS, UMR 7550, \\11 rue de
  l'Universit\'e, F-67000 Strasbourg, France}

\emailAdd{lemoine@iap.fr}
\emailAdd{kotera@iap.fr}
\emailAdd{jerome.petri@astro.unistra.fr}

\abstract{Pulsar wind nebulae (PWNe) are outstanding accelerators in
  Nature, in the sense that they accelerate electrons up to the
  radiation reaction limit. Motivated by this observation, this paper
  examines the possibility that young pulsar wind nebulae can
  accelerate ions to ultra-high energies at the termination shock of
  the pulsar wind. We consider here powerful PWNe, fed by pulsars born
  with $\sim\,$millisecond periods.  Assuming that such pulsars exist,
  at least during a few years after the birth of the neutron star, and
  that they inject ions into the wind, we find that protons could be
  accelerated up to energies of the order of the
  Greisen-Zatsepin-Kuzmin cut-off, for a fiducial rotation period
  $P\,\sim\,1\,$msec and a pulsar magnetic field
  $B_{\star}\,\sim\,10^{13}\,$G, implying a fiducial wind luminosity
  $L_{\rm p}\,\sim\,10^{45}\,$erg/s and a spin-down time $t_{\rm
    sd}\,\sim 3\times 10^7\,$s. The main limiting factor is set by
  synchrotron losses in the nebula and by the size of the termination
  shock; ions with $Z\geq 1$ may therefore be accelerated to even
  higher energies.  We derive an associated neutrino flux produced by
  interactions in the source region.  For a proton-dominated
  composition, our maximum flux lies slightly below the 5-year
  sensitivity of IceCube-86 and above the 3-year sensitivity of the
  projected Askaryan Radio Array. It might thus become detectable in
  the next decade, depending on the exact level of contribution of
  these millisecond pulsar wind nebulae to the ultra-high energy
  cosmic ray flux.  }

\keywords{Cosmic rays -- Pulsar wind nebulae}

\begin{document}
\maketitle
\flushbottom

\section{Introduction}\label{sec:introd}
The origin of the highest energy cosmic rays is a long-standing enigma
of astroparticle physics, which has withstood some fifty years of
intense experimental activity (see reviews by, e.g.,
\cite{KO11,Letessier11}). The existing data have brought in very
significant results, such as the detection of a cut-off at the
expected location of the Greisen-Zatsepin-Kuzmin (GZK)
suppression~\cite{Abbasi08,Abraham:2008ru}. Such a spectral feature,
combined with the absence of striking anisotropy in the arrival
direction of the highest energy particles, could indicate that they
originate from extragalactic sources. So far, however, no conclusive
experimental evidence points towards one or the other of the many
possible scenarios of ultra-high energy cosmic ray (UHECR) origin.

The central question in this field of research is how to accelerate
particles to these extreme energies $\sim\,10^{20}\,$eV. Among the
known particle acceleration scenarios, Fermi-type shock acceleration
plays a special role. It is rather ubiquitous, since collisionless
shock waves emerge as direct consequences of powerful
outflows. Furthermore, shock acceleration, when operative, is known to
dissipate into the supra-thermal particle population a substantial
fraction of the kinetic energy that is inflowing into the shock, of
the order of $\sim 10\%$, see e.g.~\cite{2013ApJ...771...54S}. Shock
acceleration also produces rather generically a spectrum with nearly
constant energy per decade, which allows to transfer a sizable
fraction of the energy in the ultra-high energy domain. Those are
noticeable features in the context of the origin of ultra-high energy
cosmic rays, because one indeed needs to extract a large fraction of
the source energy in order to match the cosmic ray flux above
$10^{19}\,$eV, e.g. $E_{\rm UHECR}\,\sim\, 10^{53}\,\dot
n_{-9}^{-1}\,$erg per transient source of ocurrence rate $\dot
n=10^{-9}\dot
n_{-9}\,$Mpc$^{-3}$yr$^{-1}$~\cite{Katz09}\footnote{Throughout the
  paper, quantities are noted $Q_x\equiv Q/10^x$ in cgs units, unless
  specified otherwise.}.

Particle acceleration to ultra-high energies in collisionless shock
waves has been proposed in a number of scenarios~\cite{Bykov12},
e.g. in gamma-ray
bursts~\cite{Waxman95,V95,Gialis04,W01,2010ApJ...724.1366D}, in
blazars~\cite{RB93,2010ApJ...724.1366D} or in radio-galaxy
jets~\cite{RB93}.  The possibility of accelerating ultra-high energy
cosmic rays at the termination shock of pulsar winds has received so
far little attention, except for Ref.~\cite{GA99}, which has stressed
the large energy gain associated to the first shock crossings. The
present paper thus proposes a critical discussion of this issue. Let
us recall here that the ultra-relativistic collisionless shock front
separates the (inner) cold fast magnetized pulsar wind from the pulsar
wind nebula, which itself is bounded by a shock propagating into the
supernova remnant; this nebula thus contains the hot shocked wind
material and hot shocked supernova remnant material,
e.g.~\cite{Chevalier77,Chevalier92,Gaensler05}. \\

The main motivation of the present study comes from the realization
that the Crab nebula represents so far the most efficient particle
accelerator known to us, since the observation of a synchrotron
spectrum extending up to $\sim\,m_ec^2/\alpha_{\rm em}$ ($\alpha_{\rm
  em}$ the fine structure constant) attests of the capacity of the
termination shock to accelerate electrons and positrons up to the
radiation reaction limit at the Bohm rate, meaning an acceleration
timescale $t_{\rm acc}\,\simeq\,{\cal A}\,t_{\rm g}$ in terms of the
gyro-time $t_{\rm g}$, with ${\cal A}\,\sim\,1$, expressed here in the
comoving blast frame, e.g.~\cite{Atoyan96}.  Furthermore, it is
generally admitted that the highest energy pairs have been shock
accelerated through a Fermi process, because their spectral index
$s\,\simeq\,2.2$ is remarkably similar to the predictions of
relativistic shock acceleration for isotropic scattering, see
e.g. ~\cite{BO98,Kirk00,Achterberg01,LP03,Keshet05}. In this sense, it
is natural to try to extend this result to the acceleration of
ions. Of course, pulsar winds are usually modeled as pair winds, hence
one central assumption of the present work is that such winds may also
inject ions, see the discussion in Ref.~\cite{Arons03}, and see also
~\cite{Hoshino92,Gallant94,2004ApJ...603..669S} which propose to
interpret the morphological features of the Crab Nebula through the
coupling between ion and pair dynamics at the termination shock.

The confinement energy of cosmic rays in Crab-like nebulae is
nevertheless quite low, being of the order of $E_{\rm
  conf}\,\simeq\,3\times10^{17}\,B_{-3.5}R_{18.5}\,Z\,$eV; here, $B$
and $R$ indicate respectively the magnetic field and the size of the
blast. We will thus be interested in more powerful pulsar wind
nebulae, able to confine particles up to higher energies. As we show
in the following, young pulsars born with periods
$P\,\sim\,10^{-3}\,$s do fulfill this criterion; moreover, their huge
rotational energy reservoir $E_{\rm
  rot}\,\simeq\,10^{52}\,P_{-3}^2\,$erg is also highly beneficial for
producing a substantial flux of
UHECRs~\cite{Venkatesan97,Blasi00,Fang12,Fang13}.

The release of this tremendous rotational energy into the surrounding
supernova ejecta should modify its radiative properties. In
particular, such objects could lead to ultra-luminous supernovae
lasting for months to years, with distinctive bright gamma-ray and
X-ray
counterparts~\cite{Kasen10,Woosley10,Dessart12,KPO13,Metzger14}. These
scenarios could provide an explanation to some of the observed
ultra-luminous supernovae \cite{Quimby12}, that would then constitute
an indirect probe of the existence of pulsars born with millisecond
periods. Interestingly, it has been suggested that the Crab pulsar
itself was born with a $5\,$msec period on the basis of the
observation of the large number of radio-emitting pairs in the
nebula~\cite{Atoyan99}.

Engine-driven supernovae, or trans-relativistic supernovae, whose
characteristic high speed ejecta is believed to be powered by some
internal source such as a magnetized neutron star, have also been
considered as potential sources of ultra-high energy cosmic
rays~\cite{Wang07,2012ApJ...746...40L,2011NatCo...2E.175C}. However,
in those scenarios, acceleration is argued to take place in the outer
fastest parts of the mildly relativistic external shock which
propagates in the wind of the progenitor star. Our present discussion
proposes another view of these objects, in which particle acceleration
to ultra-high energies takes place well inside the remnant, at the
ultra-relativistic shock that is running up the pulsar wind.

Finally, young pulsar winds themselves have also been considered as
potential sites of UHECR acceleration, mainly through the electric
field associated to the rotating magnetic dipole, e.g.~
\cite{Rees74,Venkatesan97,Blasi00,Bednarek02,Arons03,Fang12,Fang13}.
We stress that the present scenario is wholly different in terms of
acceleration physics. In particular, pulsar magnetospheric and wind
physics do not play any role in our scenario, beyond controlling the
spin-down time of the engine, while it directly sets the maximum
energy that might be reached in those wind acceleration models.  In
the present work, particle acceleration takes place in two steps: in a
first step, the initial Poynting flux of the pulsar wind is assumed to
be dissipated down to near equipartition with ions and electrons
before or around the termination shock; this is a generic assumption,
motivated by observational results on known pulsar wind nebula,
e.g. ~\cite{Kennel84,Kennel84b}; the injected high energy ions are
then accelerated at the termination shock, up to a maximal energy
determined by energy losses and escape considerations.

The lay-out of this paper is as follows. In Sec.~\ref{sec:pwn}, we
discuss the dynamics and the radiative properties of the nebula, which
limit the acceleration through radiative losses and escape, then we
discuss the maximal energy as a function of the various parameters. In
Sec.~\ref{sec:disc}, we discuss the neutrino signal associated to the
acceleration of ultra-high energy cosmic rays as well as the
dependence of our results on the model assumptions and parameters. We
provide a summary of our results and conclusions in
Sec.~\ref{sec:conc}.

\section{Particle acceleration in young powerful PWNe}\label{sec:pwn}

The physics of young PWNe is controlled by the amount of energy
injected by the pulsar wind, of luminosity $L_{\rm w}(t)$, and the
velocity-density profile of the supernova ejecta which confines the
nebula. Detailed morphological studies of the Crab nebula, along with
numerical MHD simulations~\cite{Komissarov04,Porth13} indicate that
the geometry is mostly axisymmetric. It is however possible to
reproduce the main characteristics of a young pulsar wind nebula with
a spherically symmetric picture, in particular the location of the
termination shock radius and the size of the nebula, e.g.,
\cite{Rees74,Kennel84}. Given the other astrophysical uncertainties
described below, this suffices for our purposes and, in the following,
we assume spherical symmetry.

Throughout this study, neutron stars have a moment of inertia
$I_\star$ with fiducial value $10^{45}\,$g$\,$cm$^2$, instantaneous
and initial rotation velocities $\Omega$ and $\Omega_{\rm i}$
(corresponding initial period $P_{\rm i}=2\pi/\Omega_{\rm i}$), radius
$R_\star$ and dipole magnetic field $B_\star$; these should not be
confused with magnetic field strengths and spatial scales of the
nebula.

The pulsar rotational energy reservoir amounts to $E_{\rm rot} =
{I_\star\Omega_{\rm i}^2}/{2} \sim 2.0\times 10^{52}\,{\rm erg}\, I_{\star,45}
P_{\rm i,-3}^{-2}$. The
wind luminosity decreases as 
\begin{equation}
L_{\rm w}(t)= L_{\rm p}/(1+t/t_{\rm  sd})^{(n+1)/(n-1)}\ ,
 \end{equation}
 in terms of the braking index $n$ (defined by
 $\dot\Omega\,\propto\,\Omega^n$) and spin-down time $t_{\rm sd}$,
 with initial luminosity $L_{\rm p}\,\simeq\,E_{\rm rot}/t_{\rm
   sd}\,\simeq\,0.64\times10^{45}
 P_{-3}^{-4}B_{\star,13}^2R_{\star,6}^6\,$erg/s. For magneto-dipole
 losses in the vacuum, $n=3$, while observations rather indicate
 $n\,\sim2-2.5$. Nevertheless, we will be interested in the structure
 of the nebula at time $t_{\rm sd}$, at which a substantial fraction
 of the rotational energy has been output into the nebula; the braking
 index controls the later evolutionary stages, therefore it will not
 impact significantly our results. We thus adopt $n=3$ for simplicity
 in what follows.

The spin-down timescale is then given by
\begin{equation}\label{eq:tp}
t_{\rm sd} \,\simeq\, \frac{9I_{\star}c^3}{8B_\star^2R_\star^6\Omega_{\rm i}^2} \sim
3.1\times 10^{7}\,{\rm  s}\,I_{\star,45}B_{\star,13}^{-2}R_{\star,6}^{-6}P_{\rm i,-3}^2\ .
\end{equation}
For convenience, we indicate our results in terms of $B_\star$,
$R_\star$ and $P$, as well as in terms of $L_{\rm p}$ and $t_{\rm
  sd}$.

We concentrate mainly on pulsars with magnetic fields
$B\,\sim\,10^{12}-10^{13}\,$G and not on magnetars (with $B\gtrsim
10^{14}\,$G). As Eq.~(\ref{eq:tp}) indicates, magnetars spin down on a
timescale much shorter than a year, and at the early times when the
highest energy particles are accelerated, the density of the
surrounding supernova ejecta does not allow their escape
\cite{Fang12}. A magnetar scenario thus require the disruption of the
supernova envelope by the wind \cite{Arons03} or that particles escape
through a region punctured by a jet, like in a strongly magnetized
proto-magnetar scenario discussed by \cite{Metzger11}. Gravitational
wave losses are negligible for pulsars with $B\ll 10^{16}\,$G, hence
we neglect them, see e.g. ~\cite{Arons03}.

\subsection{General input from the Crab nebula}\label{subsection:crab}

The extrapolation of the phenomenology of the Crab pulsar wind nebula
to the young PWNe that we are interested in is by no means trivial,
because it involves an increase by some six orders of magnitude in
luminosity, and it is hampered by three unsolved issues: the
so-called $\sigma-$problem, the physics of particle
acceleration at the termination shock of pulsar winds and the origin
of radio emitting electrons in PWNe. Let us recall briefly these
issues in order to motivate our model of the nebula.

The magnetization parameter $\sigma$ relates the Poynting flux to the
matter energy flux; in the comoving wind frame, for a cold plasma of
rest mass energy density $nmc^2$, it is defined by $\sigma\,\equiv\,
B^2/\left(4\pi nmc^2\right)$. For a mixed pair-ion composition of
respective densities $\kappa n_e$ and $n_{\rm i}$,
$nmc^2\,\equiv\,n_{\rm i} m_{\rm i} c^2 + 2\kappa n_e m_e c^2$,
$\kappa$ defining the multiplicity factor for pairs achieved through
pair cascade in the magnetosphere.
  
Observationally, the $\sigma-$problem results from the difficulty in
reconciling the large value $\sigma_0$ of the magnetization parameter
at the pulsar light cylinder with that inferred downstream of the
termination shock ($\sigma_{\rm PWN}$) through a leptonic model of the
emission~\cite{Rees74,Kennel84,Kennel84b,Atoyan96,DelZanna04,Komissarov04}. A
generic estimate for $\sigma_0$ in the Crab nebula is
$\sigma_0\,\sim\,{\cal O}(10^6)$, assuming that pairs are injected at
the Goldreich-Julian rate (recalled further below), with a low energy,
at the base of the wind, with multiplicity $\kappa\,\sim\, 10^4$. In
contrast, models of pulsar wind nebulae and their comparison to
observations rather suggest $\sigma_{\rm
  PWN}\,\sim\,10^{-3}-10^{-1}$~\cite{Kennel84,Kennel84b,Gelfand09,Bucciantini11,2014AN....335..318G,2014MNRAS.443..138M},
although more recent three-dimensional MHD simulations suggest that
values as large as $\sigma_{\rm PWN}\,\sim\,{\cal O}(1)$ could
reproduce the morphological data for the Crab
nebula~\cite{Porth13,Komissarov13}.  From a more theoretical
perspective, this $\sigma-$problem characterizes the difficulty of
pushing cold MHD winds to large Lorentz factors through the Poynting
flux,
e.g.~\cite{Coroniti90,Chiueh98,Contopoulos02,Kirk09,Lyubarsky10}.
Indeed, in a radially expanding MHD wind, $\sigma$ should remain
constant from close to the light-cylinder up to the termination shock.

How particle acceleration takes place in the Crab nebula is another
puzzle. Although the termination shock offers an obvious site for
particle acceleration, and the high energy spectral index
$s\,\simeq\,2.2$ of the reconstructed electron distribution ${\rm
  d}n/{\rm d}\gamma\,\propto\,\gamma^{-s}$ conforms well to the
expectations of a relativistic Fermi process with isotropic
scattering~\cite{BO98,Kirk00,Achterberg01,LP03,Keshet05}, so far our
understanding of particle acceleration rather suggests that the Fermi
process should be inefficient in mildly-magnetized -- namely for a
magnetization $\sigma\,\gtrsim\,10^{-4}$ -- ultra-relativistic shocks,
see ~\cite{LP10,LP11} for an analytical discussion and
~\cite{2013ApJ...771...54S} for simulations. To summarize such
discussions briefly, Fermi acceleration can take place at {\em ideal}
-- meaning planar and steady -- relativistic shock waves only if
intense small-scale turbulence has been excited in the shock
vicinity\footnote{small-scale means here $\lambda_{\delta
    B}\,<\,r_{\rm g}$, with $r_{\rm g}$ the gyroradius of accelerated
  particles}~\cite{LPR06}. Such small-scale turbulence may in
principle be excited by streaming instabilities between the
supra-thermal particles and the background unshocked plasma in the
shock precursor. At the termination shock of pulsar winds, this may
come through a current-driven instability if $\sigma\,\lesssim
10^{-2}$~\cite{LPGP13}, or through the synchrotron maser instability,
if $\sigma\,\gtrsim\,0.1$~\cite{Hoshino92,Gallant94}. Nevertheless,
the finite magnetization, even if $\sigma_{\rm PWN}$ is as small as
$10^{-3}$, should prevent acceleration to very high energies, because
the efficiency of scattering in small-scale turbulence relatively to
the gyration in the background field decreases in inverse proportion
to the particle energy~\cite{Pelletier09,LP10}. As discussed in these
references, this implies a maximum energy beyond which scattering
(hence acceleration) becomes ineffective. This maximum energy has been
observed in recent PIC simulations ~\cite{2013ApJ...771...54S}. Its
exact value is not of importance for the present discussion; it
suffices to say that it scales as $\sigma^{-1/2}$ so that, at mildly
magnetized shock waves with $\sigma\,\gtrsim\,10^{-4}$, Fermi
acceleration should not be able to accelerate the particles to the
energies observed, {\it in ideal conditions}.

Efficient dissipation of the magnetic field in the nebula, is probably
the key to resolving both the $\sigma-$problem and the issue of
particle acceleration. Regarding the former, recent 3D MHD
simulations, which account for dissipative effects inside the nebula,
alleviate somewhat the question of conversion of Poynting flux in the
wind by allowing for values $\sigma_{\rm PWN}\,\sim\,{\cal
  O}(1)$~\cite{Porth13,Komissarov13}. Although the exact mechanism of
dissipation remains open to debate, reconnection in the current sheet
separating the ``stripes" of opposite magnetic polarity (the ``striped
wind" \cite{Michel82,Michel94,Coroniti90}), upstream of the
termination shock, has long been discussed as a possible way of
converting part of the Poynting flux, e.g.~\cite{Coroniti90,Kirk03}.
Dissipation around the termination shock may also support efficient
particle acceleration in two ways: by seeding large scale turbulence,
causally disconnected from the upstream magnetic topology, in which
case particle acceleration could proceed unimpeded~\cite{LPR06}; or,
by providing an extra mechanism of particle acceleration, which would
feed into the Fermi process at higher energies, e.g. through the
dissipation of MHD waves~\cite{BS73}, or
reconnection~\cite{Lyubarsky03,Petri07,Sironi11}. Furthermore, the MHD
simulations of Ref.~\cite{Camus09} reveal that the termination shock
is unsteady and corrugated, leading to the excitation of mildly
relativistic turbulence immediately downstream. Finally, we also note
that the high energy spectral index $s=2.2$ is typical of a
relativistic Fermi process in a mildly or weakly magnetized shock,
whereas magnetized shock waves lead to a weaker compression ratio,
hence a softer spectrum~\cite{Kirk00}. This, again, argues in the
favor of substantial dissipation around the termination shock of the
wind.

One can argue further in the favor of dissipation inside the pulsar wind
nebula, as follows: ad absurdum, one could not construct a stationary
model with a super-fast magnetosonic wind in the absence of
dissipation~\cite{Kennel84}, as the shock crossing conditions at the
termination shock would then lead to a ultra-relativistic bulk
velocity for the shocked wind material; however, any unsteady solution
is bound to populate the nebula with magnetized turbulence, with a
fast magnetosonic velocity close to $c$, which would lead to particle
acceleration on a fast timescale, hence to efficient dissipation.

To summarize, both observational and theoretical arguments indicate
that efficient dissipation of the magnetic field and the tapping of
this energy into kinetic energy are relevant processes at the pulsar
wind termination shock.

The last issue concerns the origin of radio-emitting electrons in the
Crab nebula, which are about $10^2$ more numerous than the optical to
$X$-ray emitting electrons~\cite{Kirk09}. In terms of multiplicity,
the radio emission suggests a pair multiplicity
$\kappa\,\sim\,10^6$, well above the theoretical expectations
$\kappa\,\sim\,10^2-10^4$ \citep{Hibschman01,Timokhin10}, which are in
much better agreement with the multiplicity associated to higher
energy electrons.  Two generic, diverging interpretations are usually
given: either the multiplicity indeed reaches values
$\kappa\,\sim\,10^6$, in which case the pulsar spin-down power divided
by the total kinetic energy leads to a rather low value $\gamma_{\rm
  w}\,\lesssim\,10^2-10^3$ for the Lorentz factor of the wind at the
termination shock~\cite{Lyubarsky03}, or $\kappa\,\sim\,10^4$ and, for
one interpretation, the radio emitting low energy electrons were
injected at an earlier stage of the
nebula~\cite{Kennel84b,Atoyan96,Atoyan99}. Interestingly, the latter
interpretation requires that the Crab pulsar was born with a period of
order $\sim\,5\,$msec, i.e. quite close to the range of values that we
are interested in~\cite{Atoyan99}.

Let us note that the present discussion implicitly assumes values
$\kappa\,\ll\,10^6$, as if $\kappa$ were as high as $10^6$, the ions
could carry only a tiny fraction of the wind energy and it would
become difficult to match the cosmic-ray flux at the highest energies;
this issue is discussed in Sec.~\ref{sec:disc}.

In spite of the above unknowns, the fact is that the Crab accelerates
electron-positron pairs efficiently, up to the radiation reaction
limit, with a high energy index very similar to that expected in a
relativistic Fermi process, and this remains our main motivation to
discuss the possibility of pushing ions to ultra-high energies. In
what follows, we build a one-zone model of the synchrotron nebula in
order to quantify the various losses that limit the
acceleration of particles to ultra-high energies.

Following the above discussion, we assume that the wind energy is
efficiently dissipated into random particle energies inside the
nebula, i.e. $\sigma_{\rm PWN}\,\lesssim\,1$; hereafter, $\sigma_{\rm
  PWN}$ is understood as the average magnetization parameter inside
the nebula, after dissipation has taken place. This dissipation can
either take place upstream of the termination shock, in which case the
shock itself transforms the ordered kinetic energy into random
particle energies; or, it can take place at or downstream of the
termination shock, through one of the various processes discussed
above. We also assume that the termination shock is strong, which
implies super-fast magnetosonic velocities of the wind; this, however,
is not a stringent requirement on wind physics, since it only requires
a wind Lorentz factor at the termination shock $\gamma_{\rm
  w}\,\gtrsim\, \sigma_{\rm i}^{1/3}\,\sim\,\gamma_{\rm
  diss.}^{1/3}\,\sim\,10^3$ for our fiducial parameters, see the
definition of $\gamma_{\rm diss.}$ in Eq.~(\ref{eq:erec}) below, and
$\sigma_{\rm i}$ corresponds to the initial magnetization of the
wind~\cite{Kirk09}.

\subsection{The millisecond nebula structure}\label{sec:nebula}

The structure of the nebula can be approximated by analytical
solutions at times $t\,\lesssim\,t_{\rm sd}$, when $L_{\rm w}$ is
approximately constant~\cite{Reynolds84}. The pulsar wind nebula
radius can then be written
\begin{eqnarray}
R_{\rm PWN} &\,=\,& \left(\frac{125}{99}\frac{\beta_{\rm
    PWN}^3c^3L_0}{M_0}\right)^{1/5}t^{6/5} \qquad
\left(t_{\rm c}\,\gtrsim\,t\right)\label{eq:RC1}\\ &\,=\,&\left(\frac{8}{15}\frac{L_0}{M_0}\right)^{1/2}t^{3/2}
\qquad \left(t_{\rm sd} \,\gtrsim\,t\,\gtrsim\,t_{\rm c}\right)\ ,\label{eq:RC2}
\end{eqnarray}
in terms of the time $t_{\rm c}$ at which the external shock of the
PWN has swept up the mass of the supernova ejecta, $M_0$, which we
assume of constant density (with a rapidly declining density profile
beyond) and in terms of the wind power $L_0$; $\beta_{\rm PWN}$
represents the velocity of the pulsar wind nebula in the source rest
frame. For parameters of interest, one finds that $t_{\rm
  sd}\,\gtrsim\,t_{\rm c}$, since $t_{\rm c}\,\sim\,10^5\,{\rm
  s}\,L_{45}^{-1}M_{\rm ej,34}v_{\rm ej83}^2$ for a core mass $M_{\rm
  ej}\,=\,5M_{\rm ej,34}M_\odot$ and an ejecta velocity $v_{\rm
  ej}\,=1000\,v_{\rm ej,8}\,$km/s~\cite{Reynolds84,KPO13}, hence we
will use mostly Eq.~(\ref{eq:RC2}) in the following.

At times $t\,\gtrsim\,t_{\rm sd}$, the pulsar input into the nebula
decreases rapidly. For this phase, we then assume that the blast
evolves in free expansion, meaning ~\cite{KPO13}
\begin{equation}
R_{\rm PWN}(t)\,=\,R_{\rm PWM}(t_{\rm sd})\frac{t}{t_{\rm
    sd}}\label{eq:fexp}
\end{equation}

Of course, we are mostly interested in the PWN structure at the time
$t_{\rm sd}$, since it corresponds to the time of maximum energy
injection into the nebula. The temporal scalings of $R_{\rm PWN}$ at
times $t\,\ll\,t_{\rm sd}$ and $t\,\gg\,t_{\rm sd}$ thus do not play a
crucial role in the forthcoming analysis, but they help in
understanding how the various quantities evolve in time.

The above estimates neglect the interaction of the blast with the
outer shocked region of the supernova, i.e. the forward and reverse
shocks associated with the interaction with the circumstellar
medium. In more typical, less powerful ($L_{\rm
  p}\,\lesssim\,10^{39}\,$erg/s) pulsar wind nebulae, the interaction
with the reverse shock takes place on timescales of thousands of years
and this leads to the compression of the nebula ~\cite{Gelfand09}. In
the present case, the interaction takes place shortly after $t_{\rm
  c}$, i.e. shortly after the PWN external shock into the ejecta has
swept up the inner core of the remnant. However, the pulsar energy
output $E_{\rm rot}\,\sim\,2\times 10^{52}\,P_{-3}^2\,$ergs dominates
the kinetic energy of the outer blast ($E_{\rm
  ej}\,\sim\,10^{51}\,$ergs), therefore the nebula will dominate the
dynamics. We neglect this interaction phase; again, it should not
modify appreciably the values of $R_{\rm PWN}$ that we derive at time
$t_{\rm sd}$, which is our prime objective here.

The above analytical solutions also fail when radiative cooling of the
blast becomes important. As we show in the following, the latter
possiblity is to be considered, because the electrons cool through
synchrotron faster than a dynamical timescale, contrary to what
happens in PWNe such as the Crab. Therefore, if dissipation of the
Poynting flux is efficient, and if ions represent a modest part of the
energy budget, most of the wind luminosity input into the nebula is
actually lost into radiation. In order to account for this effect, we
use an improved version of Eqs.~(\ref{eq:RC1}),(\ref{eq:RC2}), in
which $L_0=(1-\eta_{\rm rad})L_{\rm p}$ represents the actual power
deposited into the nebula, $\eta_{\rm rad}=1-\eta_B-\eta_{\rm i}$
representing the fraction of luminosity converted into radiation
through pair cooling ($\eta_B$: fraction of energy in the magnetic
field, $\eta_{\rm i}$: fraction of energy in ions, in the nebula).
These approximations are used to provide analytical estimates of the
various quantities characterizing the nebula at time $t_{\rm sd}$.

We complement these estimates with a detailed numerical integration of
the following system:
\begin{eqnarray}
\dot R_{\rm PWN}&\,=\,&\beta_{\rm PWN}c\label{eq:rb}\\
\dot R_{\rm es}&\,=\,&\beta_{\rm es}c\label{eq:res}\\
\dot U_{\rm sw}&\,=\,& \left(\beta_{\rm
    w}-\beta_{\rm ts}\right)L_{\rm w} - P_{\rm em} - 4\pi R_{\rm
  PWN}^2 \beta_{\rm PWN}c\,p_{\rm PWN},\label{eq:uswd}\\
\dot M_{\rm se}&\,=\,& \left(\beta_{\rm
    es}-\beta_{\rm ej}\right) 4\pi r_{\rm es}^2 \rho_{\rm ej}
c,\label{eq:msed}\\
\dot U_{\rm se}&\,=\,& \left(\beta_{\rm
    es}-\beta_{\rm ej}\right) 4\pi r_{\rm es}^2 \rho_{\rm ej}
c^3 + 4\pi R_{\rm PWN}^2\beta_{\rm PWN}c\,p_{\rm PWN} .\label{eq:used}
\end{eqnarray}
All quantities are defined in the source rest frame; they are as
follows: $R_{\rm PWN}$ corresponds to the radius of the contact
discontinuity, interpreted as the size of the nebula; $R_{\rm es}$
represents the location of the outer shock of the nebula, propagating
in the supernova remnant; to a very good approximation, $R_{\rm
  es}\,\simeq\,R_{\rm PWN}$ (thin-shell approximation); $\beta_{\rm
  es}$ consequently represents the velocity of this outer shock while
$\beta_{\rm ts}$ denotes the velocity of the termination shock;
$U_{\rm sw}$ represents the energy contained in the shocked wind
region, beneath the contact discontinuity; $P_{\rm em}$ represents the
power lost through radiation; since the electrons cool faster than an
expansion timescale (see below), one can write $P_{\rm
  em}\,=\,\left(1-\eta_B-\eta_{\rm i}\right)\left(\beta_{\rm
  w}-\beta_{\rm ts}\right)L_{\rm w}$, in terms of the fraction of
power injected into the nebula in magnetic field ($\eta_B$) and ions
($\eta_{\rm i}$). $p_{\rm PWN}$ represents the pressure inside the
nebula, which can be well approximated by $U_{\rm sw}/(4\pi R_{\rm
  PWN}^3)$~\cite{Bucciantini11}; the term associated to $p_{\rm PWN}$
consequently represents adiabatic losses for $U_{\rm sw}$; $M_{\rm
  se}$ denotes the mass accumulated in the shocked ejecta region,
between the contact discontinuity and the outer shock; $\beta_{\rm
  ej}$ corresponds to the velocity of the supernova remnant ejecta, in
the source frame; finally, $U_{\rm se}$ denotes the energy contained
in the shocked ejecta region. To a good approximation, $U_{\rm
  se}-M_{\rm se}c^2\,\simeq\,M_{\rm se}\beta_{\rm PWN}^2c^2/2$ since the
ejecta is non-relativistic.

These equations can be obtained by integrating the equations of
particle current density and energy-momentum conservation over the
spatial variables, between the boundaries of interest. This procedure
introduces the brackets $\left(\beta_{\rm w}-\beta_{\rm ts}\right)$
for $\dot U_{\rm sw}$ and $\left(\beta_{\rm es}-\beta_{\rm ej}\right)$
for $\dot U_{\rm se}$, which correspond to the fact that the boundaries of
the shocked wind and shocked ejecta are delimited by the moving shock
waves. The velocity of the termination shock in the source frame
depends non-trivially on the degree of magnetization of the shock; for
$\sigma\,\ll\,1$, however, $\beta_{\rm ts}\,\ll\,\beta_{\rm
  w}$~\cite{Kennel84} and $\beta_{\rm w}\,\simeq\,1$, therefore we
approximate $\beta_{\rm w}-\beta_{\rm ts}\,\simeq\,1$. For the outer
shock, assuming it is strong, non-radiative and non-relativistic, one
has $\beta_{\rm es}-\beta_{\rm ej}\,\simeq\, 4(\beta_{\rm
  PWN}-\beta_{\rm ej})/3$. This closes the system.

\begin{figure}[t]
\centering
\includegraphics[width=0.48\textwidth]{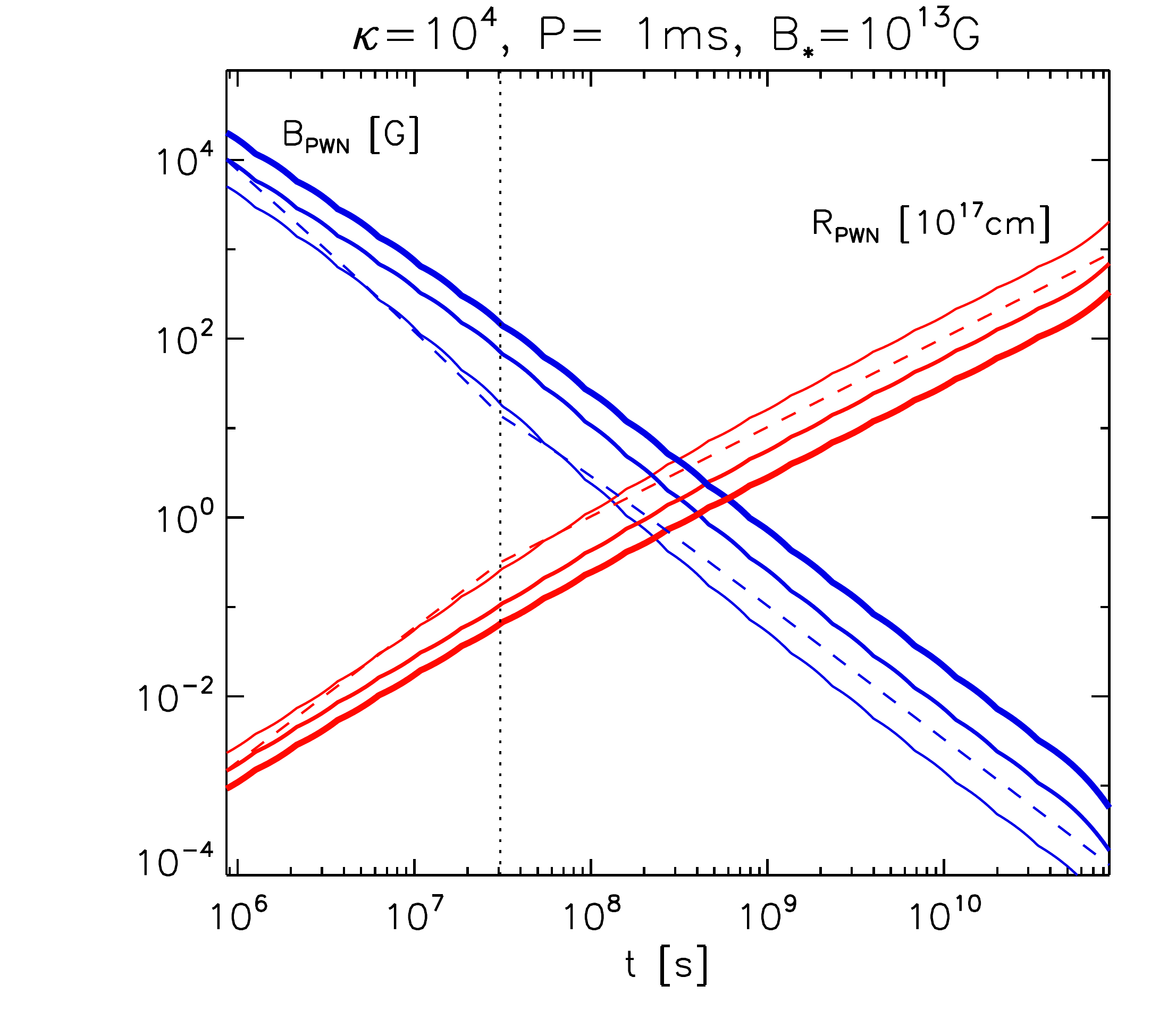} 
\includegraphics[width=0.48\textwidth]{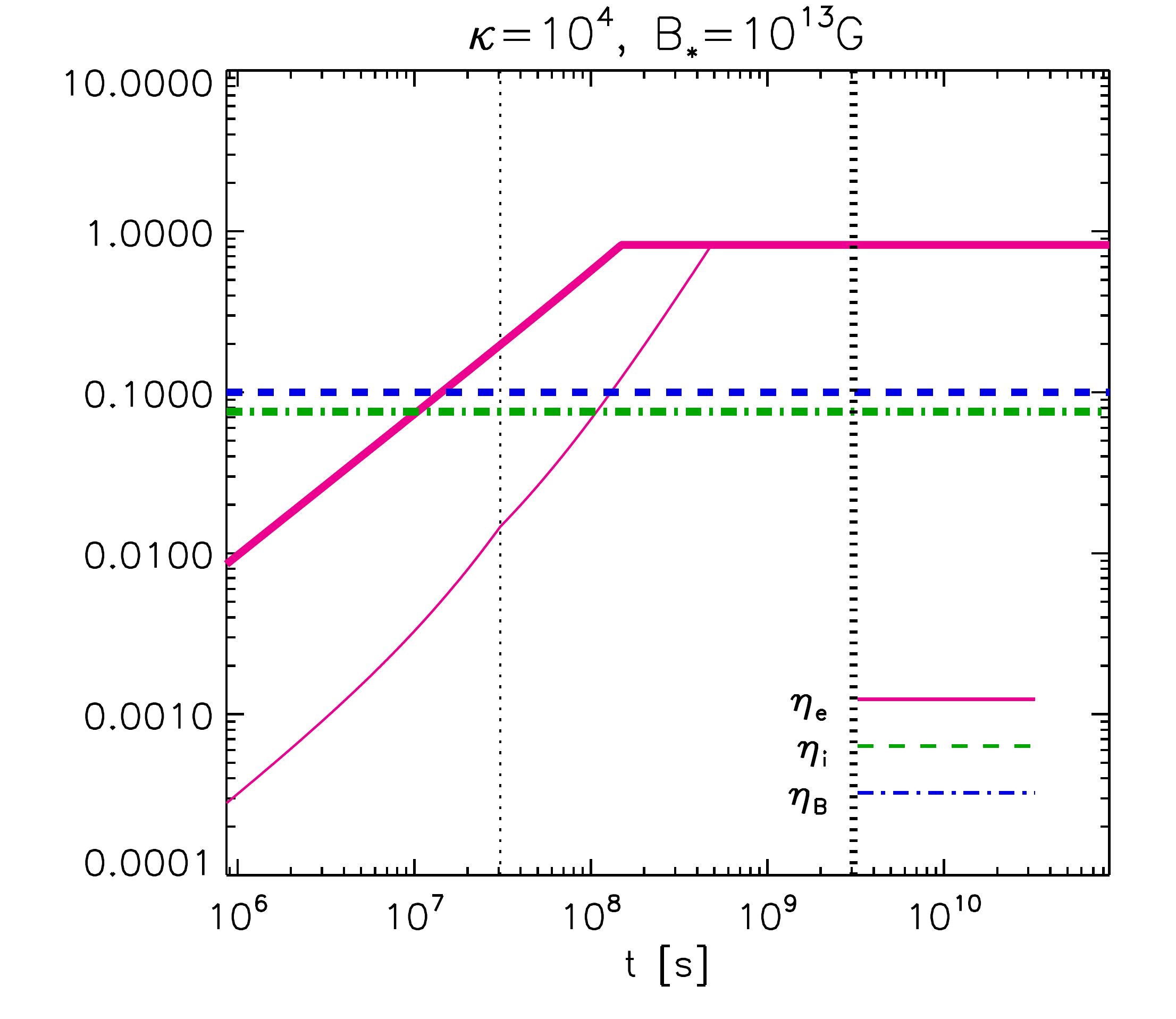} 
\caption{Evolution in time of pulsar wind dynamical quantities, for
  pulsar dipole magnetic field $B_{\rm \star, 13}=1$, leptonic
  multiplicity $\kappa_{4}=1$ and assuming $\eta_{\rm B}=0.1$. {\it
    Left:} Case of a pulsar with initial rotation period
  $P_{-3}=1$. Red increasing lines: radius of the nebula (obtained by
  integrating numerically
  Eqs.~\ref{eq:rb}$-$\ref{eq:used}). Increasing thickness for
  $\eta_{\rm rad}=0,0.9,0.99$. The dashed lines represent the
  analytical adiabatic case (Eq.~\ref{eq:RPWN}). Blue decreasing
  lines: corresponding mean magnetic field in the pulsar wind nebula
  $B_{\rm PWN}$ (Eq.~\ref{eq:BPWN}). {\it Right:} Fraction of the
  pulsar luminosity dissipated into magnetic energy in the nebula
  ($\eta_B$, blue dashed), to leptons ($\eta_e$, pink solid) and to
  ions ($\eta_{\rm i}$, assuming protons, green dot-dashed). Initial
  rotation periods $P_{-3}=1$ (thin lines), and $P_{-3}=10$ (thick
  lines). The vertical dotted line indicates the spin-down timescale
  $t_{\rm sd}$. \label{fig:PWN}}
\end{figure}

In order to evaluate the dynamics of the nebula, we assume that the
supernova ejecta consists of a core mass $5\,M_\odot$ of constant
density. An analytical estimate of the size of the pulsar wind nebula
$R_{\rm PWN}$ is then given by Eqs.~(\ref{eq:RC2}) and
  (\ref{eq:fexp}):
\begin{eqnarray}
R_{\rm PWN}&\,\simeq\,& 4.1\times 10^{16}\,L_{\rm p,45}^{1/2}t_{\rm
  sd,7.5}^{3/2}\check t^{3/2}\hat t\,\,{\rm
  cm}\nonumber\\ &\,\simeq\,& 3.2\times 10^{16}\,
P_{-3}B_{\star,13}^{-2}R_{\star,6}^{-6}I_{\star,45}^{3/2}\,\check
t^{3/2}\hat t\,\, {\rm cm}\ .\label{eq:RPWN}
\end{eqnarray}
The quantities $\check t\,\equiv\,{\rm min}\left(1,t/t_{\rm
  sd}\right)$ and $\hat t\,\equiv\,{\rm max}\left(1,t/t_{\rm
  sd}\right)$ indicate the scaling of these values at times
respectively short and long of $t_{\rm sd}$, obtained respectively
through Eq.~(\ref{eq:RC2}) and
Eq.~(\ref{eq:fexp}). Figure~\ref{fig:PWN} presents the evolution in
time of pulsar wind dynamical quantities ($R_{\rm PWN}$ and $B_{\rm
  PWN}$) calculated analytically and by numerical integration. The
prefactors match the numerical evaluation shown in Fig.~\ref{fig:PWN}
for $\eta_{\rm rad}\lesssim 0.9$. The scaling departs slightly from
the $t^{3/2}$ (resp. $t$) behaviour at short (resp. late) times
compared to $t_{\rm sd}$, but we neglect this difference in the
following. Radiative nebulae, in which $\eta_{\rm rad}$ is closer to
unity, tend to be more compact; this difference can be read off
Fig.~\ref{fig:PWN} and inserted in the relations that follow. For the
sake of simplicity, we assume an adiabatic case in the following, i.e.
$1-\eta_{\rm rad}\sim 1$.

At this stage, it may be useful to make contact with known pulsar wind
nebulae, such as the Crab: its radius $R_{\rm PWN}$ is about $1\,$pc,
i.e. about a hundred times larger than the above. In the Crab nebula,
the radius of the termination shock is estimated to be
$\sim\,0.1\,$pc~\cite{Kennel84}, while in the present case, the
termination shock is located close to the contact discontinuity, see
Sec.~\ref{sec:conf}, making the above nebula not only more compact,
but also much thinner.

The mean magnetic field in the nebula $B_{\rm PWN}$ is then obtained
as follows. Recall that $\eta_B$ corresponds to the magnetic fraction
of the energy actually injected into the nebula after a proper account
of dissipation, i.e. $\eta_B\,=\,\sigma_{\rm PWN}/(1+\sigma_{\rm
  PWN})$; one thus has
\begin{eqnarray}\label{eq:BPWN}
B_{\rm PWN} &\,=\,& \left(\frac{6\eta_{\rm B} \int_0^{t} L_{\rm
    w}(t'){\rm d}t'}{R_{\rm PWN}^3}
\right)^{1/2}\nonumber\\ &\,\simeq\,& 14 \,\eta_{B,-1}^{1/2}L_{\rm p,45}^{-1/4}t_{\rm
  sd,7.5}^{-7/4}\check t^{-7/4}\hat
  t^{-3/2}\,\,{\rm G}\nonumber\\ &\,\simeq\,&
12\,P_{-3}^{-5/2}\eta_{B,-1}^{1/2}I_{\star,45}^{-7/4}B_{\star,13}^3R_{\star,6}^9\,\check
t^{-7/4}\hat t^{-3/2}\,\, {\rm G}\ .
\end{eqnarray}
The numerical values are obtained by plugging into the first equation
the temporal scalings of $L_{\rm w}$ and $R_{\rm PWN}$ obtained
previously. Analytical and numerical estimates agree for the adiabatic
case at the spin-down time, see Fig.~\ref{fig:PWN}. The magnetic field
strength is of course much larger than that seen in more standard
PWNe~\cite{Gelfand09,2014JHEAp...1...31T,Bucciantini11}, as a result
of the larger input energy and of the younger age, which implies a
more compact nebula.

The right panel of Fig.~\ref{fig:PWN} depicts the evolution of the
distribution of the energy fractions $\eta_{\rm i}$, $\eta_e$ and
$\eta_{\rm B}$, discussed below.\\

\subsection{Energy injected into particles in the nebula}\label{sec:injg}
As argued in Section~\ref{subsection:crab}, we assume efficient
dissipation of the initial Poynting flux, into random particle
energy. In the absence of energy losses, this conversion implies that
particles (electron-positron pairs or ions) acquire a typical Lorentz
factor
\begin{eqnarray}
  \gamma_{\rm diss.}&\,\simeq\,&\frac{1}{1+\sigma_{\rm
      PWN}}\frac{L_{\rm w}}{\dot N m
    c^2}\nonumber\\ 
&\,\simeq\,&2.2\times 10^9\frac{1-\eta_{\rm B}}{1 + x_{\rm
      i}}\kappa_4^{-1}L_{\rm
    p,45}^{1/2}\hat t^{-1}\nonumber\\ 
&\,\simeq\,& 1.8\times 10^9\frac{1-\eta_{\rm B}}{1 + x_{\rm
      i}}\,\kappa_4^{-1}P_{-3}^{-2}B_{\star,13}R_{\star,6}^3\hat t^{-1}\ .\label{eq:erec}
\end{eqnarray}
This Lorentz factor $\gamma_{\rm diss.}$ can also be written as:
$\gamma_{\rm diss.}=\gamma_{\rm w}(1+\sigma_{\rm ts})/(1+\sigma_{\rm
  PWN})$ in terms of $\sigma_{\rm ts}$, the magnetization of the flow
short of the termination shock. The particle rest mass power injected
into the nebula is written here: $\dot N mc^2 \,\equiv\, 2\kappa\,\dot
N_{\rm GJ} m_e c^2(1+x_i)$, with $\dot N_{\rm GJ}= e^{-1}\sqrt{L_{\rm
    w}c}$ the Goldreich-Julian rate~\cite{Goldreich69};
$x_{\rm i}$ is the ratio of the power injected into ions, relatively
to that injected into pairs: if ions of charge $Z_{\rm i}$ are
injected at a rate $\dot N_{\rm GJ}/Z_{\rm i}$, $x_{\rm i}\,\equiv\,
m_{\rm i}/\left(2Z_{\rm i}\kappa m_e\right)$, so that $x_{\rm
  i}\,\lesssim\,1$ for $\kappa\,\gtrsim\,10^3$.

Up to radiation reaction effects, heating through dissipation proceeds
equally for pairs and ions, meaning that both acquire a same Lorentz
factor. This is guaranteed for all dissipation processes mentioned
earlier. Among others, this implies that, notwithstanding radiation
reaction effects, the ratio of the energy injected into pairs to that
injected into ions is conserved in the dissipative processes.

At the present time, one cannot predict what kind of ions the pulsar
would output. We therefore remain general and consider ions of mass
number $A_{\rm i}$, charge $Z_{\rm i}$. Note that the composition of
the highest energy cosmic rays is not very well-known either: while
the Pierre Auger Collaboration reports a light composition at
$10^{19}\,$eV, transiting to an intermediate composition at higher
energies~\cite{2014PhRvD..90l2006A}, the results of the Telescope
Array experiment seemingly point towards a light
composition~\cite{2015APh....64...49A}, even though the depths of
shower maximum of both experiments appear compatible, see
~\cite{2015arXiv150307540A}.

Actually, energy losses may limit the typical Lorentz factor
$\gamma_e$ of electron-positron pairs to the minimum of $\gamma_{\rm
  diss.}$ and the radiation reaction limiting Lorentz factor
$\gamma_{\rm e-loss}$. Of course, $\gamma_e$ cannot be lower than the
  Lorentz factor of the wind at the termination shock, $\gamma_{\rm
    w}$, which is unknown. We assume that $\gamma_{\rm w}\,<\,{\rm
    min}\left(\gamma_{\rm diss.},\gamma_{\rm e-loss}\right)$. This is
  not a strong assumption, since the latter two Lorentz factors are
  quite large.

Equating the synchrotron cooling time with the gyration time of the
particle, the radiation reaction limiting Lorentz factor $\gamma_{\rm
  e-loss}$ is given by the usual formula
\begin{eqnarray}
  \gamma_{\rm e-loss} &\,=\,&\frac{3}{2}\frac{m_e c^2}{e^{3/2} B_{\rm
      PWN}^{1/2}}\nonumber\\ 
&\,\simeq\,& 3\times 10^7 \eta_{B,-1}^{-1/4}L_{\rm
    p,45}^{1/8}t_{\rm sd,7.5}^{7/8}\check t^{7/8}\hat
  t^{9/8}\nonumber\\ 
&\,\simeq\,&
  3\times10^7\,\eta_{B,-1}^{-1/4}P_{-3}^{5/4}I_{\star,45}^{7/8}B_{\star,13}^{-3/2}R_{\star,6}^{-9/2}\check
  t^{7/8}\hat t^{9/8}\ .
\end{eqnarray}

The synchrotron power of ions of Lorentz factor $\gamma$, atomic
number $Z_i$ and mass number $A_i$ scales as $P_{\rm syn}\,=\,(Z_{\rm
  i}^4/A_{\rm i}^2)(m_e/m_p)^2(4/3)\sigma_{\rm T}U_Bc\gamma^2\beta^2$,
therefore the radiation reaction Lorentz factor for ions is a factor
$A_{\rm i}Z_{\rm i}^{-3/2}m_p/m_e$ larger than $\gamma_{\rm
  e-loss}$. For ions, radiation reaction therefore does not limit the
efficiency of dissipation.

Accounting for dissipation and radiation reaction limitations, one can
thus write the fractions of energy $\eta_e$ and $\eta_i$ carried by
the electrons and the ions inside the nebula as:
\begin{equation}
\eta_e\,=\, \frac{1-\eta_B}{1+x_{\rm i}}\,{\rm
  min}\left(1,\frac{\gamma_{\rm e-loss}}{\gamma_{\rm
    diss.}}\right)\ ,\quad
\eta_i\,=\, \frac{(1-\eta_B)x_{\rm i}}{1+x_{\rm i}}\ .\label{eq:etae}
\end{equation}
The quantity $\eta_e$ is understood as characterizing the energy
injected in pairs in the nebula, after dissipation/acceleration
processes, but before synchrotron cooling has taken place.  Clearly,
for the above fiducial parameters, $\eta_e\,\ll\,1$ at time $t_{\rm
  sd}$, meaning that most of the energy dissipated into the electrons
has been radiated at the radiation reaction limit, producing photons
of energy $\sim\,50\,$MeV. This radiation does not contribute to the
radiation losses of ultra-high energy ions but it may lead to a
specific signature of dissipation processes in such young PWNe. In
contrast, $\gamma_{\rm e-loss}\,\gg\,\gamma_{\rm diss.}$ in the Crab
nebula, so that the electrons can take away most of the dissipated
energy without losing it to radiation.

In such compact PWNe, the electrons cool through synchrotron radiation
on a timescale that is much shorter than a dynamical timescale, down
to non-relativistic velocities, since the cooling Lorentz factor is
given by
\begin{eqnarray}
\gamma_{\rm c}&\,\simeq\,&\frac{6\pi m_e c^2 \beta_{\rm
    PWN}}{\sigma_{\rm T} B_{\rm PWN}^2 R_{\rm
    PWN}}\nonumber\\ &\,\sim\,&\, t_{\rm sd,7.5}^2\beta_{\rm
  PWN}\eta_{B,-1}^{-1}\check t^2\hat
  t^2\nonumber\\ &\,\sim\,&\,\eta_{B,-1}^{-1}P_{-3}^{4}\beta_{\rm
  PWN}B_{\star,13}^{-4}R_{\star,6}^{-12}I_{45}^{2}\check
t^2\hat t^2\ .
\end{eqnarray}
This represents a major difference with respect to the case of the
Crab nebula, for which $\gamma_{\rm c}\,\gg\,1$, so that most
electrons do not cool on an expansion timescale, due to the smaller
amount of energy injected into the wind and to the larger size of the
nebula.

The above allows us to characterize the spectral energy distribution
(SED) of the nebula; in particular, the low-frequency spectral
luminosity is represented by
\begin{equation}
L_{\nu,\rm syn}\,\simeq\, \eta_e L_{\rm w}
\,\left(\frac{\epsilon}{\epsilon_{e}}\right)^{1/2}\quad
\left(\epsilon_{\rm
  c}<\epsilon<\epsilon_{e}\right) \label{eq:sLnu}
\end{equation}
with $\epsilon_{\rm c}=h\nu_{\rm c}$ and $\epsilon_{e}=h\nu_{\rm e}$
in terms of the synchrotron peak frequencies associated to Lorentz
factors $\gamma_{\rm c}$ and $\gamma_e=\min(\gamma_{\rm
  e-loss},\gamma_{\rm diss.})$. In Fig.~\ref{fig:PWN}, we plot the
time evolution of the various quantities that characterize the nebula,
i.e. the mean magnetic field, the mean nebular radius and the
fractions of energy $\eta_e$, $\eta_B$ and $\eta_{\rm i}$.

\subsection{Ion acceleration}

After their injection through the termination shock, the ions are
energized through dissipative processes up to $\gamma_{\rm diss.}$,
then shock accelerated to the maximal Lorentz factor $\gamma_{\rm
  max}$ that we seek to determine here. This latter Lorentz factor is
given, as usual, by the competition between shock acceleration, escape
from the PWN and energy losses in the synchrotron nebula.

We assume here that acceleration proceeds at the Bohm rate, an
assumption that is motivated and supported by the two following
remarks. The first is of a more empirical nature as it follows from
the observation that the Crab nebula does accelerate electron-positron
pairs up to the radiation reaction limit. If the acceleration
timescale is written $t_{\rm acc}\,=\,{\cal A}t_{\rm g}$, then the
comparison of $t_{\rm acc}$ with the synchrotron loss timescales leads
to an upper bound on the maximum photon energy,
$\epsilon_\gamma\,\lesssim\,{\cal A}^{-1}\,m_e c^2/\alpha_{\rm
  em}$. The fact that the synchrotron spectrum extends up to
$\sim60\,$MeV or so in the Crab nebula indicates that ${\cal
  A}\,\sim\,1$.

The second line of argument originates from our theoretical
understanding of particle acceleration at relativistic shock waves; to
put it briefly, ${\cal A}\,\sim\,1$ if some large scale turbulence,
seeded downstream of the termination shock by dissipative processes,
mediates the scattering process at the termination shock. In order to
see this, one must recall that supra-thermal particles probe a short
length scale of order $\sim\,r_{\rm g}$ behind a relativistic oblique
shock, before returning to the shock or being advected
away~\cite{LPR06}. Therefore, if some turbulence is transmitted from
upstream to downstream through the shock, the shock crossing
conditions imply the continuity of the magnetic field lines through
the shock, which then prevent repeated Fermi cycles, unless most of
the turbulent power lie on short length
scales~\cite{Niemiec06,LPR06,Pelletier09,LP10}. If, however, the
turbulence is seeded downstream of the shock, which requires
additional dissipative processes as in the present case, then this
continuity is broken, hence Fermi cycles can take place, as modelled
in test-particle Monte Carlo simulations,
e.g.~\cite{BO98,Kirk00,Achterberg01,LP03}. The downstream and upstream
residence timescales are then both of order $r_{\rm g}$ in the shock
rest frame, so that the acceleration timescale corresponds to ${\cal
  A}\,\sim\,{\cal O}(1)$.

\subsubsection{Confinement at the shock and in the nebula}\label{sec:conf}
Confinement in the nebula itself leads to a maximal Lorentz factor:
\begin{eqnarray}
\gamma_{\rm conf}&\,\sim\,& \frac{Z_{\rm i}e B_{\rm PWN} R_{\rm
    PWN}}{m_{\rm i} c^2}\nonumber\\ &\,\simeq\,&
1.5\times10^{11}\,\frac{Z_{\rm i}}{A_{\rm i}}\,
\eta_{B,-1}^{1/2}L_{\rm p,45}^{1/4}t_{\rm sd,7.5}^{-1/4}\check
t^{-1/4}\hat t^{-1/2}\\ &\,\simeq\,&1.4 \times 10^{11}\,\frac{Z_{\rm
    i}}{A_{\rm i}}\,\eta_{B,-1}^{1/2} P_{-3}^{-3/2}
I_{\star,45}^{-1/4}B_{\star,13}R_{\star,6}^3\check t^{-1/4}\hat
t^{-1/2}\ .\label{eq:gconf}
\end{eqnarray}
The large value of $\gamma_{\rm conf}$ confirms that young fast
pulsars input enough power into the nebula to confine particles up to
ultra-high energies, a non-trivial result in itself.

As a matter of fact, the finite size of the termination shock limits
the maximum acceleration energy to a factor $\sim 2$ below the above
confinement energy, because the shock radius $r_{\rm
  ts}\,\sim\,\alpha_{\rm ts}\,R_{\rm PWN}$ with $\alpha_{\rm
  ts}\,\sim\,0.4$. This value is derived from the numerical
simulations discussed in Sec.~\ref{sec:nebula}, but it can be
understood as follows. In the source rest frame, one can write the
velocity of the termination shock to first order in $1/\gamma_{\rm w}$
as follows,
\begin{equation}
\beta_{\rm ts}\,\simeq\,\frac{1-3\beta_2}{-3+\beta_2}\label{eq:bts}
\end{equation}
where $\beta_2$ denotes the velocity of the shocked wind, immediately
downstream of the termination shock.  This equation assumes
hydrodynamic jump conditions at the shock to simplify the analysis. It
is essentially a rewriting of the shock velocity in a frame in which
the downstream is moving at velocity $\beta_{\rm PWN}$; the expression
in the downstream frame is obtained by $\beta_2\,\rightarrow\,0$,
which implies $\beta_{\rm ts}\,\rightarrow\,-1/3$, as expected for a
strong ultra-relativistic hydrodynamic shock
wave~\cite{1976PhFl...19.1130B}. One commonly assumes that the flow
velocity then evolves as a function of radius according to
$\beta\,\propto\,r^{-2}$, see ~\cite{Kennel84,Kennel84b} for a
detailed discussion; this implies that the blast velocity $\beta_{\rm
  PWN}\,\simeq\,\beta_2(R_{\rm PWN}/r_{\rm ts})^2$. Since $R_{\rm
  PWN}/r_{\rm ts}\,\simeq\,\beta_{\rm PWN}/\beta_{\rm ts}$, one can
solve $r_{\rm ts}/R_{\rm PWN}$ as a function of $\beta_{\rm PWN}$,
using the above in conjunction with Eq.~(\ref{eq:bts}). One finds
\begin{equation}
r_{\rm ts}\,\simeq\,\left[\sqrt{3\beta_{\rm b}}\,+{\cal O}(\beta_{\rm PWN}^2)\right]R_{\rm PWN}\ .
\end{equation}
Therefore, $\alpha_{\rm ts}\,\simeq\,\sqrt{3\beta_{\rm
    PWN}}\,\sim\,0.4$ for a typical velocity $\beta_{\rm
  PWN}\,\sim\,0.05$ at $t_{\rm sd}$ (a value checked in numerical
calculations).

At a gyroradius $r_{\rm g}$ a factor of order unity to a few above
$r_{\rm ts}$, the particle feels the shock curvature and it gradually
decouples from the flow as its size exceeds the size of the
accelerator. Acceleration therefore stops at Lorentz factor
\begin{equation}
\gamma_{\rm max}\,\simeq\,\alpha_{\rm ts}\gamma_{\rm conf}\label{eq:gts}
\end{equation}

\subsubsection{Synchrotron losses}
Due to the strong magnetic field in the nebula, synchrotron losses
represent a potential limitation to the maximum energy. In the course
of acceleration, synchrotron losses limit $\gamma$ to
\begin{eqnarray}
\gamma_{\rm i,syn-acc}&\,\simeq\,& 6.2\times
10^{10}\eta_{B,-1}^{-1/4}A_{\rm i}Z_{\rm i}^{-3/2} L_{45}^{1/8}t_{\rm
  sd,7.5}^{7/8}\check t^{7/8}\hat
t^{3/4}\nonumber\\ &\,\simeq\,&6.0\times
10^{10}\eta_{B,-1}^{-1/4}A_{\rm i}Z_{\rm
  i}^{-3/2}P_{-3}^{5/4}B_{\star,13}^{-3/2}R_{\star,6}^{-9/2}I_{\star,45}^{7/8}\check t^{7/8}\hat
t^{3/4}\label{eq:syn1}
\end{eqnarray}
As discussed in Sec.~\ref{sec:injg}, this maximum Lorentz factor is
simply $\gamma_{e}$ times $A_{\rm i}Z_{\rm i}^{-3/2}m_p/m_e$. It lies
a factor of a few below the confinement energy, but it nevertheless
allows protons to be accelerated up to energies of the order of the
GZK cut-off for our fiducial parameters.

Given that $r_{\rm ts}\,\lesssim\,R_{\rm PWN}$, particles also suffer
synchrotron energy losses during their escape out of the nebula. The
corresponding cooling Lorentz factor is obtained by matching $P_{\rm
  syn}R_{\rm PWN}/c$ and the particle energy, which leads to
\begin{eqnarray}
\gamma_{\rm i,syn-esc}&\,\simeq\, & 2.6\times
10^{10}\,\eta_{B,-1}^{-1}A_{\rm i}^3Z_{\rm i}^{-4}t_{\rm
  sd,7.5}^2\check t^2\hat t^2\nonumber\\ &\,\simeq\,& 2.4\times
10^{10}\,\eta_{B,-1}^{-1}A_{\rm i}^3Z_{\rm
  i}^{-4}P_{-3}^4B_{\star,13}^{-4}R_{\star,6}^{-12}I_{\star,45}^2\check
t^2\hat t^2\label{eq:syn2}
\end{eqnarray}
As a result of its combination of powers of $t_{\rm sd}$, $\hat t$ and
$\check t$, the above equation can actually be rewritten as
$\gamma_{\rm i,syn-esc}\,\simeq\, 2.5\times
10^{10}\,\eta_{B,-1}^{-1}A_{\rm i}^3Z_{\rm i}^{-4}\left(t/{\rm
  1\,yr}\right)^2$, with a numerical prefactor entirely controlled by
the core mass of the supernova ejecta and by fundamental
constants. However, we are interested in computing the limiting
Lorentz factor at a time $t_{\rm sd}$, at which most of the rotational
energy has been output by the pulsar; for this reason, one should
actually read Eq.~(\ref{eq:syn2}) above with $\hat t=\check t =1$,
which {\it de facto} introduces a dependence on $t_{\rm sd}^2$.

This limit appears more severe than the previous ones, but with a
rather strong dependance on the spin-down time $t_{\rm sd}$, or
alternatively, on the dipole moment $\mu=B_\star R_\star^3/2$ of the
neutron star, see Eq.~(\ref{eq:tp}). Therefore, a spin-down time
larger by a factor two, or a (modest) factor $\sqrt{2}$ decrease in
$\mu$ would push this maximum energy [as well as Eq.~(\ref{eq:syn1})]
at or above $10^{20}\,$eV for protons, while the maximal energy
associated to the finite size of the termination shock at $\hat
t=\check t = 1$ would remain of the order to $10^{20}\,$eV.  Note that
the total injected energy $E\,\sim\,E_{\rm rot}\,\simeq\,L_{\rm
  w}t_{\rm sd}$ does not depend on $\mu$.  Alternatively, if one
assumes $\eta_B\,\sim\,1$, a magnetic field $B_\star\,\sim\,3\times
10^{12}\,$G would still guarantee that the limiting synchrotron loss
energy lies above the GZK energy (for protons), just as the maximum
energy associated to the finite size of the termination shock.

Of course, synchrotron energy losses are much weaker for ions with
$A_{\rm i}>1$; the corresponding maximal energy is larger than that of
protons by a factor $(A_{\rm i}/Z_{\rm i})^4\,\sim\,16$.

\subsubsection{Photohadronic losses}
Assuming that the injected ions are protons, the impact of photopion
interactions can be evaluated in the standard $\Delta-$approximation,
according to which interactions take place with photons of energy
$\epsilon_{\gamma\pi}\sim 0.3\,{\rm GeV}/\gamma_{\rm p}$, with
$\gamma_{\rm p}$ the proton Lorentz factor and cross-section
$\sigma_{\gamma\pi}\sim 0.5\,$mb. It is straightforward to check that
all protons interact with photons in the low-frequency part
$\lesssim\,\nu_e$ of the synchrotron SED, which justifies our neglect
of the high-frequency part of the synchrotron SED.  The number density
of such low-frequency synchrotron photons is\footnote{for clarity, we
  assume $1+x_{\rm i}\,\sim\,1$, $1-\eta_B\,\sim\,1$ in the
  expressions that follow}
\begin{equation}
  \epsilon_\gamma \frac{{\rm d}n_\gamma}{{\rm d}\epsilon_\gamma}\,\simeq\, \eta_e
  \left(\frac{\epsilon_\gamma}{\epsilon_{e}}\right)^{1/2}\frac{L_{\rm
      w}}{4\pi R_{\rm PWN}^2 c\epsilon_\gamma}\quad \left(\epsilon_{\rm
    c}\,<\,\epsilon_\gamma\,<\,\epsilon_e\right)
\end{equation}
so that the pion production optical depth, $\tau_{\gamma\pi}=R_{\rm
  PWN}\sigma_{\gamma\pi}\epsilon_\gamma{\rm d}n_\gamma/{\rm
  d}\epsilon_\gamma$ reads:
\begin{eqnarray}
\tau_{\gamma\pi} &\,\simeq\,& 7.9\times 10^{-3}\,\eta_e\, \gamma_{{\rm
    p},11}^{1/2}L_{\rm p,45}^{1/2}t_{\rm
  sd,7.5}^{-3/2}\check t^{-3/2}\hat t^{-3}
\nonumber\\ &\,\simeq\,& 6.6\times 10^{-3}\,\eta_e\, \gamma_{{\rm
    p},11}^{1/2}P_{-3}^{-5}B_{\star,13}^4R_{\star,6}^{12}I_{\star,45}^{-3/2}\check
t^{-3/2}\hat t^{-3}\ ,\label{eq:tpig1}
\end{eqnarray}
with $\gamma_{\rm p,11}=\gamma_{\rm p}/10^{11}$. Note that $\eta_e$
depends on time, in particular $\eta_e \,\ll\,1$ at early times, see
Eq.~(\ref{eq:etae}) and Fig.~\ref{fig:PWN}, while $\eta_e \,\sim\,1$
at late times.

\begin{figure}[tbp]
\centering
\includegraphics[width=0.48\textwidth]{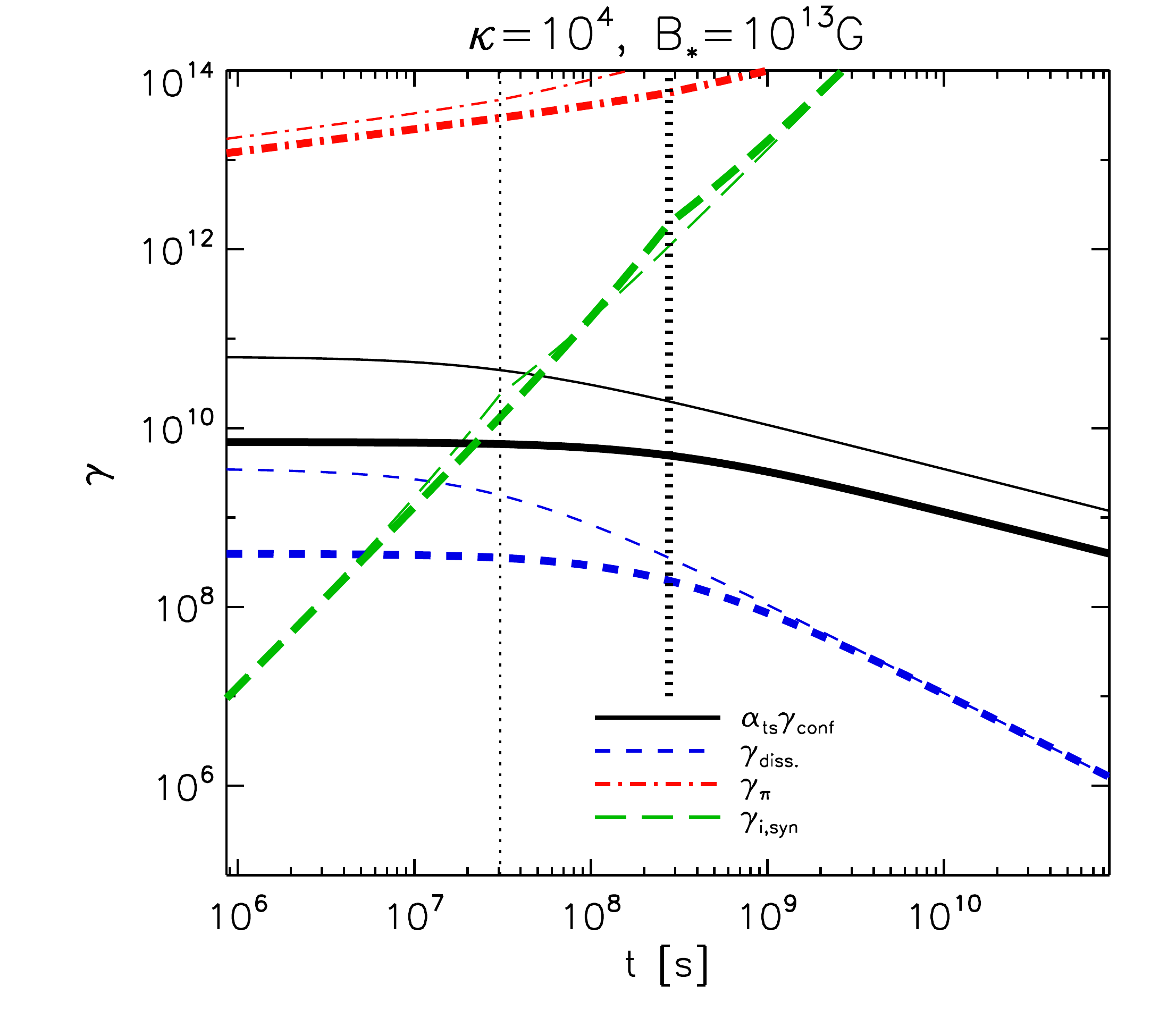} 
\includegraphics[width=0.48\textwidth]{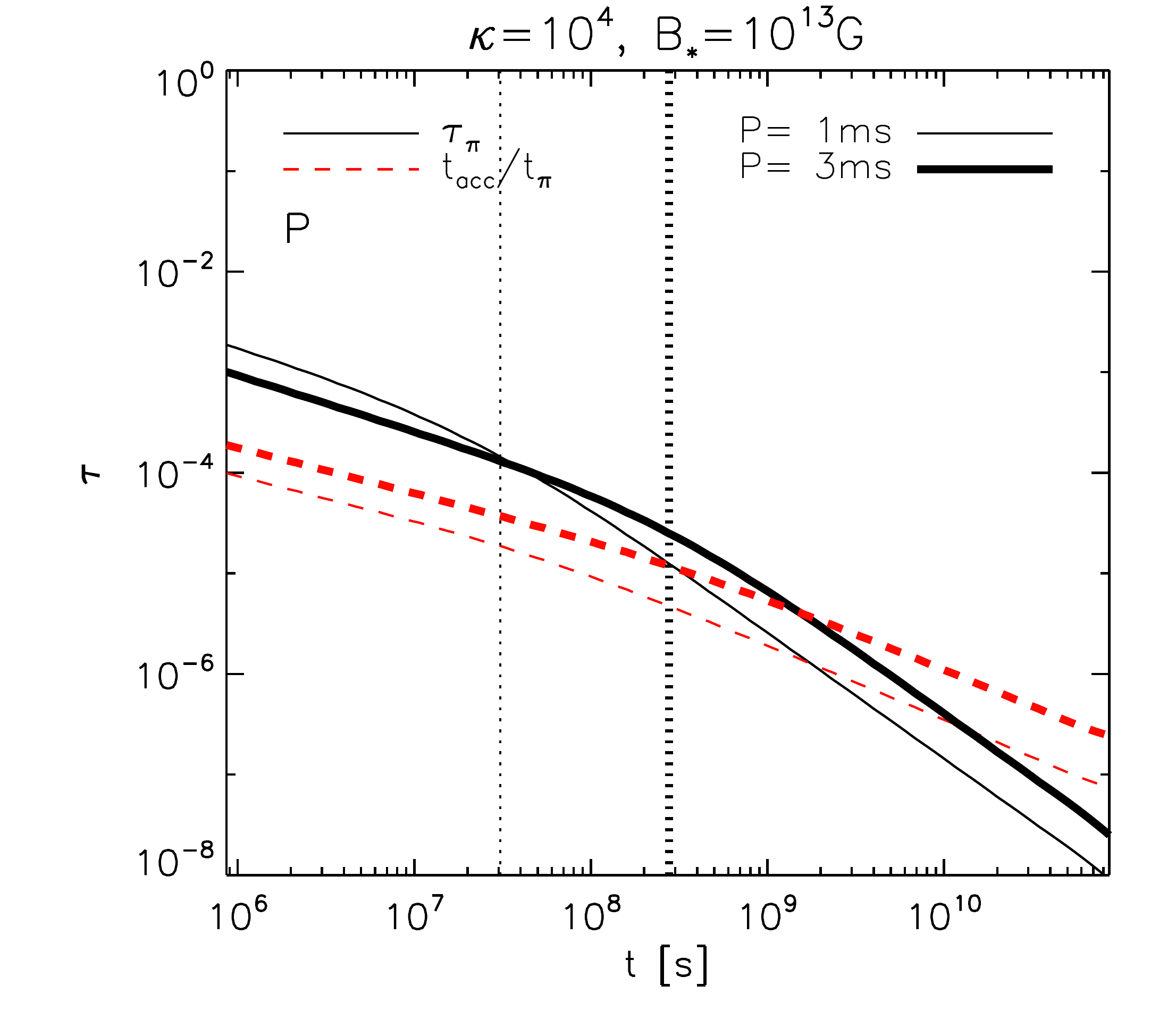} 
\caption{{\it Left:} Comparison of maximum Lorentz
  factors for proton confinement ($\gamma_{\rm conf}$, solid lines),
  acceleration ($\gamma_{\rm diss.}$, dashed lines) and energy loss by
  pion production ($\gamma_\pi$, red dot-dashed lines) and by
  synchrotron cooling ($\gamma_{\rm i,syn}$, green long-dashed lines),
  for pulsar initial rotation period $P_{-3}=1, 3$ (increasing
  thickness), dipole magnetic field $B_{\rm \star, 13}=1$, leptonic
  multiplicity $\kappa_{4}=1$, $\eta_{\rm rad}=0$ and $\eta_{\rm
    B}=0.1$. The vertical dotted line indicates the spin-down
  timescale $t_{\rm sd}$ corresponding to each rotation period
  (increasing thickness). See the text for details, in particular
  regarding the dependence of these curves on the parameters of the
  neutron star. {\it Right:} Corresponding pion production optical
  depth (black solid) and ratio of the acceleration timescale, $t_{\rm
    acc}$, to the pion production timescale, $t_{\gamma\pi}$, (red
  dashed) for a proton at Lorentz factor $10^{11}$. 
 \label{fig:gamma_many}}
\end{figure}

The ratio of the acceleration timescale to pion production timescale,
\begin{eqnarray}
  \frac{t_{\rm acc}}{t_{\rm
      \gamma\pi}}&\,\simeq\,&\frac{\gamma_p}{\gamma_{\rm
      conf}}\tau_{\gamma\pi}\ ,\label{eq:tacctpi1}
\end{eqnarray}
indicates that pion production is not a limiting factor
for acceleration to the highest energies.

The left panel of figure~\ref{fig:gamma_many} presents the time
evolution of the proton confinement Lorentz factor $\gamma_{\rm
  conf}$, dissipation Lorentz factors $\gamma_{\rm diss.}$, and
limiting Lorentz factor from synchrotron losses $\gamma_{\rm i,syn}$
and pion production interactions $\gamma_\pi$, for initial periods
$P=1,3,10\,$ms.

The evolution over time of the pion production optical depth and the
ratio of the acceleration to pion production timescales, calculated at
the at energy $\gamma_{\rm p,11}$ are shown in the right panel of 
Fig.~\ref{fig:gamma_many}. 

\begin{figure}[tbp]
\centering 
\includegraphics[width=0.48\textwidth]{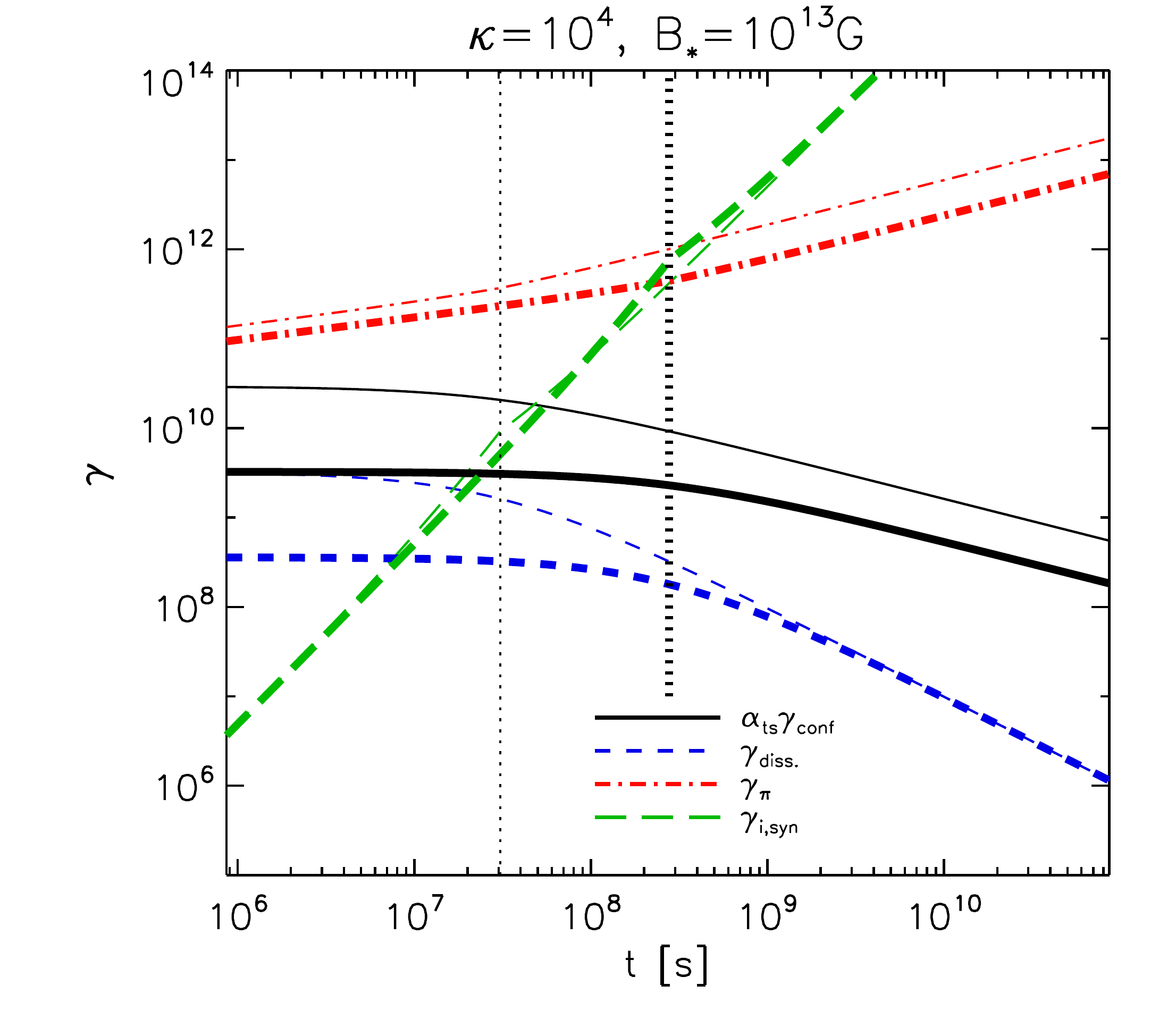} 
\includegraphics[width=0.48\textwidth]{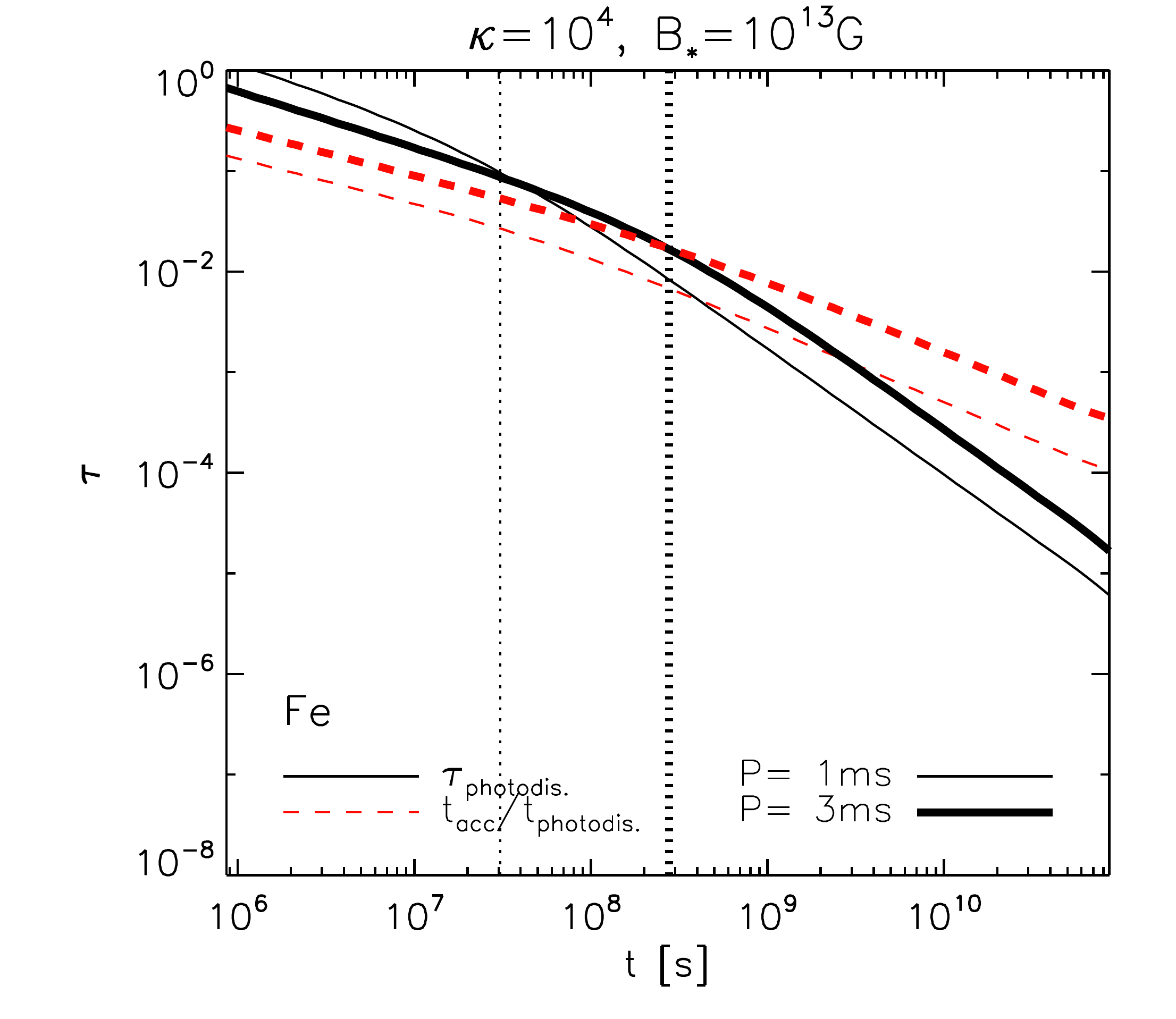} 
\caption{Same as Fig.~\ref{fig:gamma_many}, but for iron nuclei.} \label{fig:tau_Fe}
\end{figure}

The same type of calculations can be performed for heavier nuclei,
considering the Giant Dipole Resonance (GDR) as the main channel for
energy losses on the background photons. The parameters for such
interactions read: $\sigma_{A\gamma}\sim 8\times
10^{-26}\,A_{56}\,$cm$^{-2}$ for the cross-section, and $\epsilon_{\rm
  GDR}\sim 18\,A_{56}^{-0.21}\,{\rm MeV}/\gamma_{\rm i}$
\cite{SS99}. The results for iron nuclei are presented in
Fig.~\ref{fig:tau_Fe}.

All in all, Equations~(\ref{eq:gts}), (\ref{eq:syn1}), (\ref{eq:syn2})
and (\ref{eq:tacctpi1}) and the accompanying discussion thus indicate
that proton (and heavier ion) acceleration to energies of the order of
or above the GZK cut-off appears possible in PWNe with parameters
close to those chosen in this paper, namely $B_{\star,13}\,\sim\,1$,
$P_{-3}\,\sim\,1\,$ms, $\eta_B\,\sim\,0.1-1$.

\section{Discussion}\label{sec:disc}

Equations~\ref{eq:gconf}, (\ref{eq:syn1}), (\ref{eq:syn2}),
\ref{eq:tpig1} and \ref{eq:tacctpi1} summarize the confinement and
acceleration properties of young powerful pulsar wind nebulae at time
$t_{\rm sd}$, when most of the rotational energy is output into the
nebula. A third constraint on potential ultra-high energy cosmic ray
sources comes from the energy output into cosmic rays above
$10^{19}\,$eV. For a proton dominated composition, this energy output
must match $\dot \epsilon\,\simeq\, 0.5 \times
10^{44}\,$ergs/Mpc$^3$/yr once it is folded over the population of
sources with ocurrence rate $\dot n$~\cite{Katz09}.

Over $t_{\rm sd}$, the pulsar injects into ions
\begin{equation}
E_{\rm i}\,=\,\eta_{\rm i}\,L_{\rm w}\,t_{\rm sd}
\,=\,\frac{(1-\eta_B)x_{\rm i}}{(1+x_{\rm i})}L_{\rm p}t_{\rm
  sd}\ .\label{eq:Ei1}
\end{equation}
Assuming that these msec PWNe constitute a fraction $\eta_{\rm SN}$ of
the total core collapse supernovae, with ocurrence rate $\dot n_{\rm
  SN}\,\sim\,5\times 10^{-5}\,$Mpc$^{-3}$yr$^{-1}$, the normalization
to the flux of ultra-high energy cosmic rays thus implies
\begin{equation}
\eta_{\rm SN}\,\sim\,3\times 10^{-5}q_{\rm i}\,x_{\rm i}^{-1}\,L_{\rm
  p,45}^{-1}t_{\rm sd,7.5}^{-1}\ ,\label{eq:esn1}
\end{equation}
assuming here $x_{\rm i}\,\lesssim\,1$, i.e. $\kappa\,\gtrsim\,10^3$,
and $1-\eta_{\rm B}\,\sim\,1$. Recall that $x_{\rm i}\,\simeq\,0.09
(A_{\rm i}/Z_{\rm i})\kappa_4^{-1}$, indicating that these young msec
pulsars should constitute a fraction $\sim 0.03\%$ of the supernova
rate. If one rather wishes to normalize the flux at an energy
$10^{18}\,$eV with $s\sim 2$, then this fraction would go up to $\sim
1\%$.  The prefactor $q_{\rm
  i}\equiv\,(s-2)/\left\{1-\left[\gamma_{\rm max}(t_{\rm
    sd})/\gamma_{\rm diss.}(t_{\rm sd})\right]^{2-s}\right\}$ for an
injection spectral index $s$, accounts for the difference in
normalization induced by the lower cosmic-ray injection energy limit.

\subsection{Neutrino signal for a proton-dominated composition}\label{sec:nu}
The detection of neutrinos associated with hadronic and photo-hadronic
interactions of nuclei in the nebula would provide an unambiguous test
of the present scenario. Let us first consider the yield of neutrinos
through photo-hadronic interactions on the nebula SED. Since the
neutrino yield for heavy nuclei is much smaller than that for protons,
we assume in this section that $Z_{\rm i}=A_{\rm i}=1$. The neutrino
spectrum is then shaped by the accelerated proton spectrum and by the
conversion efficiency. A parent proton produces 3 neutrinos per $p-n$
conversion, which takes place with probability $1/3$ in each photopion
interaction, thus on a timescale $t_{\gamma\pi^+}=3t_{\gamma\pi}$. At
the detector, the neutrino carries a fraction
$f_\nu/(1+z)\,\simeq\,0.05/(1+z)$ of the parent proton energy,
i.e. $E_{\nu}\,\simeq\,f_\nu E_{\rm i}/(1+z)$, with $z$ the redshift
of the source.

The time-dependent neutrino spectrum emitted by one PWN at luminosity
distance $D_{\rm L}$ can be written:
\begin{equation}
E_\nu^2\Phi_\nu(E_\nu,t)\,=\, \frac{(1+z)}{4\pi D_{\rm L}^2} E_\nu^2
\frac{3}{t_{\gamma\pi^+}(E_{\rm i})}\frac{{\rm d}E_{\rm i}}{{\rm d}E_\nu}
  \frac{{\rm d}N_{\rm i}}{{\rm
      d}E_{\rm i}}\ ,\label{eq:nflux}
\end{equation}
in terms of the parent proton spectrum ${\rm d}N_{\rm i}/{\rm d}E_i$
in the source.  The latter is formally given by
\begin{equation}
 \frac{{\rm d}N_{\rm i}(t)}{{\rm d}E_{\rm i}}\,=\,\eta_{\rm
   i}\int_0^t\,q_{\rm i}(t')\frac{L_{\rm w}(t')}{E_{\rm
     diss.}^2}\,\left[\frac{E_{\rm i,0}(t')}{E_{\rm
       diss.}(t')}\right]^{-s}\,\frac{{\rm d}E_{\rm i,0}(t')}{{\rm
     d}E_{\rm i}}\,e^{-\int_{t'}^{t}{\rm d}t''/t_{\rm
     exit}\left[t'',E_{i,0}(t'')\right]}\,{\rm d}t'\ ,\label{eq:pspec}
\end{equation}
with $q_{\rm i}(t')\,=\,(s-2)/\left\{1-\left[\gamma_{\rm
    max}(t')/\gamma_{\rm diss.}(t')\right]^{2-s}\right\}$ a
normalization prefactor.  The energy $E_{\rm
  diss.}\,\equiv\,\gamma_{\rm diss.}m_pc^2$, while $E_{\rm i,0}(t')$
represents the energy of the proton at time $t'$, which shifts down to
$E_{\rm i}$ at time $t$ due to adiabatic losses. The timescale $t_{\rm
  exit}\,=\,\left(t_{\gamma\pi^+}^{-1}+t_{\rm esc}^{-1}\right)^{-1}$
represents the timescale on which protons leave the source, either
through $p-n$ conversion or through direct escape on timescale $t_{\rm
  esc}$. This solution neglects the energy loss associated with
photo-pion production; the latter is small and the probability of
exiting directly the source as a neutron significant, therefore
photo-pion production effectively acts as a loss term from the nebula.

In order to simplify the above calculation, we make the usual
approximation, e.g.~\cite{1998PhRvD..58l3005R}, that neutrino
production takes place during a timescale $t_{\rm loss}=\left[t_{\rm
    ad}^{-1}+t_{\rm esc}^{-1}+t_{\gamma\pi+}^{-1}\right]^{-1}$ and
that the proton spectrum can be described by its time average on that
timescale. Here, $t_{\rm ad}\,=\,R_{\rm PWN}/(\beta_{\rm PWN}c)$
characterizes the adiabatic loss timescale. Direct escape takes place
on timescale $t_{\rm esc}\,\simeq\, R_{\rm PWN}\left(\gamma_{\rm
  i}/\gamma_{\rm conf}\right)^{-1}$, corresponding to the assumption
of diffusive escape with a Bohm scattering timescale $\sim\,r_{\rm
  g}/c$. The average proton spectrum is
\begin{equation}
\left\langle\frac{{\rm d}N_{\rm i}(t)}{{\rm d}E_{\rm
    i}}\right\rangle\,=\, {\rm min}\left(1,\frac{t_{\rm loss}}{t_{\rm
    sd}}\right)\,
\frac{\eta_iq_iL_{\rm p}t_{\rm sd}}{E_{\rm diss.}^2}\left(\frac{E_{\rm
    i}}{E_{\rm diss.}^2}\right)^{-s}\ .
\end{equation}
Indeed, if $t_{\rm loss}\,\gg\,t_{\rm sd}$, the pulsar luminosity
function can be approximated as impulsive, $L_{\rm w}\,\sim\,t_{\rm
  sd}L_{\rm p}\delta(t-t_{\rm sd})$, while if $t_{\rm
  loss}\,\ll\,t_{\rm sd}$, the luminosity is approximately constant up
to time $t_{\rm sd}$, but at any time, the energy contained in
protons is a fraction $t_{\rm loss}/t_{\rm sd}$ of the energy injected
over time $t_{\rm sd}$. One thus derives a neutrino energy flux
\begin{equation}
E_\nu^2\Phi_\nu\,\simeq\,\frac{1+z}{4\pi D_{\rm
    L}^2}\frac{f_\nu}{t_{\gamma\pi}}\,{\rm min}\left(1,\frac{t_{\rm
    loss}}{t_{\rm sd}}\right) \eta_{\rm i}q_{\rm i}L_{\rm p}t_{\rm
  sd}\left(\frac{E_{\nu}}{E_{\nu\star}^2}\right)^{2-s}\ ,\label{eq:nuflux}
\end{equation}
with $E_{\nu\star}\,\equiv\,f_\nu E_{\rm diss.}/(1+z)$. 

The diffuse neutrino flux produced by such PWNe can be evaluated as
follows. Writing $\dot n_{\rm s}$ the ocurrence rate per comoving
volume element, the effective density at any time is ${\rm
  max}\left(t_{\rm loss},t_{\rm sd}\right)\,\dot n_{\rm s}$, since
${\rm max}\left(t_{\rm loss},t_{\rm sd}\right)$ indicates the
effective duration of neutrino emission. The diffuse energy flux then
reads
\begin{equation}
E_\nu^2j_\nu\,=\,\frac{c}{4\pi}\int_{0}^{+\infty}\,\frac{{\rm
    d}z}{H(z)(1+z)}\,\dot n_{\rm s}\frac{t_{\rm
    loss}}{t_{\gamma\pi}}f_\nu\,\eta_{\rm i}q_{\rm i}L_{\rm p}t_{\rm
  sd}\left(\frac{E_{\nu}}{E_{\nu\star}^2}\right)^{2-s}\ .\label{eq:diffnuflux}
\end{equation}
Note that $\dot n_{\rm s}$ may contain a redshift dependence, and
that $t_{\rm loss}$ and $t_{\gamma\pi}$ depend on the (source
rest-frame) energy $(1+z)E_\nu$.

In order to evaluate the neutrino flux that results from $pp$
interactions in the nebula environment, we consider only the
interactions that arise as the cosmic rays cross the supernova
remnant, given the very low density of particles within the nebula
itself. As discussed in Ref.~\cite{Fang12}, the optical depth to $pp$
interactions during the crossing of a $10M_\odot$ supernova remnant
can be written 
\begin{equation}
\tau_{pp}\,\simeq\, \frac{0.3\,{\rm yr}}{t_{\rm esc}}\ .
\end{equation}
The $pp$ interactions only take place whenever $t_{\rm esc}\,<\,t_{\rm
  ad}$ and $t_{\rm esc}\,<\,t_{\gamma\pi^+}$. Given the small optical
depth to photo-pion production, the latter condition is always
satisfied; the former amounts to $\gamma\,>\,\beta_{\rm
  PWN}\gamma_{\rm conf}$, therefore the $pp$ neutrino signal only
concerns the highest energy range. It is then straightforward to
evaluate the diffuse $pp$ neutrino flux from
Eq.~(\ref{eq:diffnuflux}), making the substitution
$t_{\rm loss}/t_{\gamma\pi}\,\rightarrow\,{\rm min}(\tau_{pp},1)
ct_{\rm esc}/R_{\rm PWN}$.

\begin{figure}[tbp]
\centering
\includegraphics[width=0.48\textwidth]{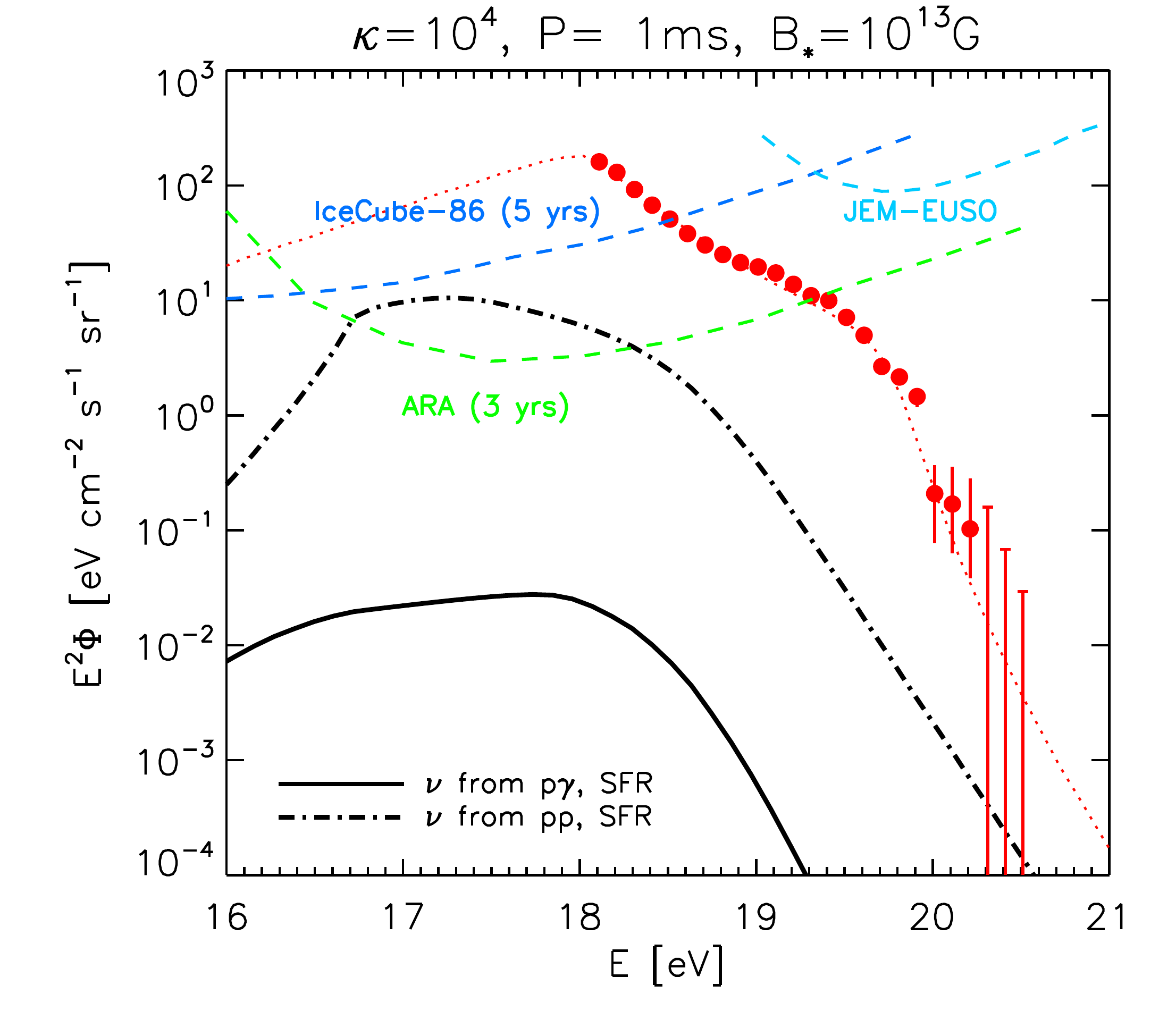} 
\includegraphics[width=0.48\textwidth]{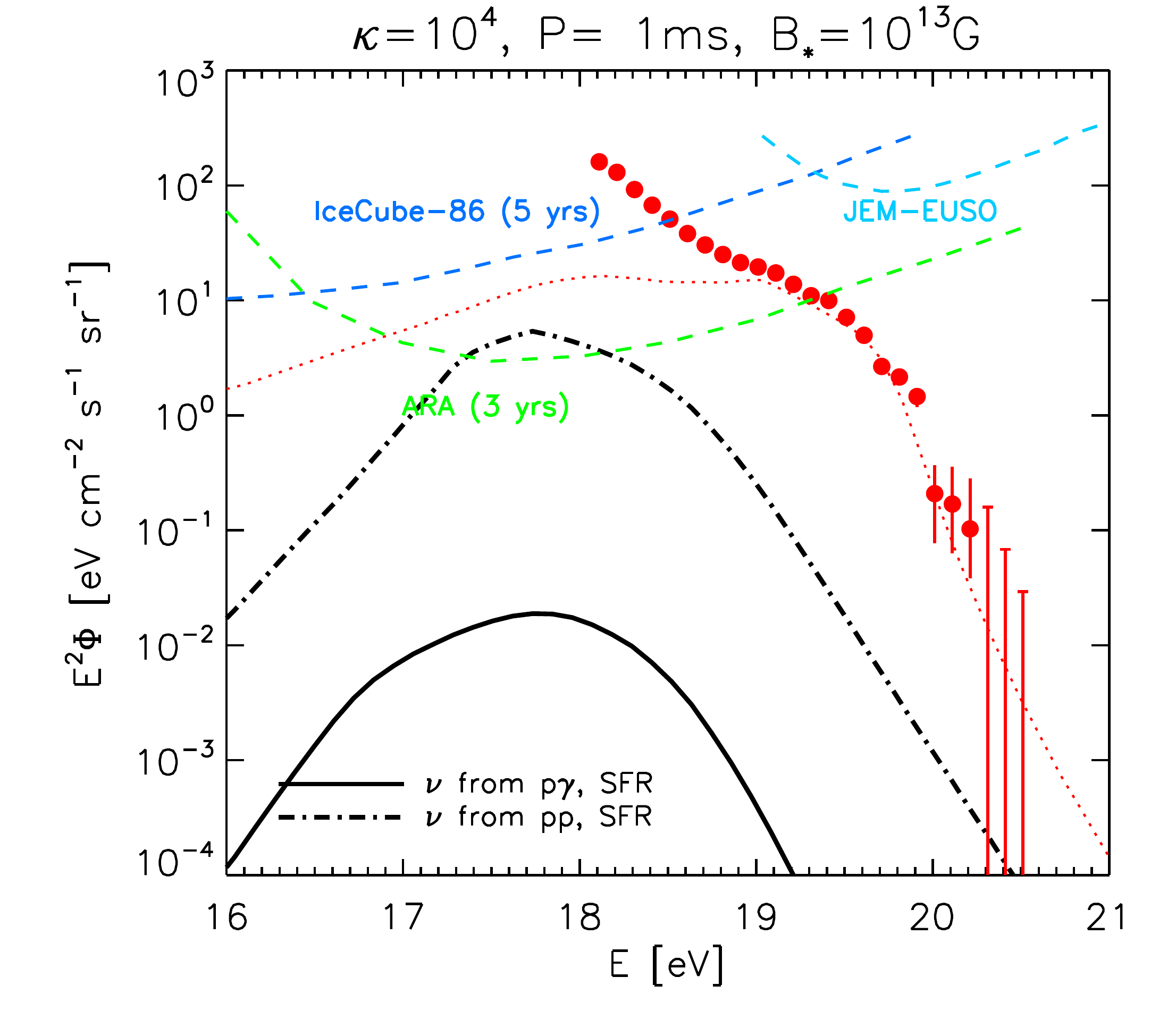} 
\caption{Neutrino spectra produced via $pp$ (dot-dashed) and $p\gamma$
  (solid) interactions for a population of pulsars with initial
  rotation period $P_{-3}=1$, dipole magnetic field $B_{\rm \star,
    13}=1$, leptonic multiplicity $\kappa_{4}=1$, $\eta_{\rm rad}=0$,
  $\eta_{\rm B}=0.1$; left panel: cosmic ray flux normalized to the
  data at $>10^{18}\,$eV; right panel: cosmic ray flux normalized to
  the data at $>10^{19}\,$eV . The sensitivity of IceCube-86 for 5
  years \citep{Abbasi:2011zx}, JEM-EUSO \citep{Adams:2012tt}, and
  ARA-37 for 3 years \citep{Allison12} are overplotted. In red dotted
  lines, the fit to the UHECR spectrum measured by the Auger
  Observatory \citep{ThePierreAuger:2013eja}, in red circles.
\label{fig:neutrinos}}
\end{figure}

Figure~\ref{fig:neutrinos} presents the neutrino spectrum produced by
protons accelerated with a spectral index of $s=2.2$, for a population
of pulsars with identical parameters, with initial rotation period
$P_{-3}=1$, dipole magnetic field $B_{\rm \star, 13}=1$, leptonic
multiplicity $\kappa_{4}=1$, $\eta_{\rm rad}=0$, $\eta_{\rm
  B}=0.1$. The birth rate of these sources are assumed to have an
occurrence rate scaled to the star formation rate (SFR) with $\dot
n_{\rm s} = 800\,$Gpc$^{-3}$\,yr$^{-1}$ at $z=0$ (i.e. $\simeq
1-2\,$\% of the supernova rate) in the left panel, which match the
cosmic-ray flux at energies as low as $10^{18}\,$eV, in order to
provide a maximum neutrino flux for this scenario. The right panel
shows the expected neutrino flux if the cosmic-ray flux is matched at
energies $>10^{19}\,$eV, corresponding to a more reasonable occurence
rate $n_{\rm s}\,=\,200\,$Gpc$^{-3}$\,yr$^{-1}$ at $z=0$. The
calculation of the cosmic-ray spectrum considers energy losses during
propagation in the intergalactic medium.

The maximum neutrino flux produced by $pp$ interactions lies slightly
below the 5-year sensitivity of IceCube-86 and above the 3-year
sensitivity of the projected Askaryan Radio Array (ARA). It might thus
become detectable in the next decade, depending on the exact level of
contribution of these msec PWNe to the ultra-high energy cosmic ray
flux. Note that if the primaries were heavier nuclei and not protons,
these estimates would be reduced by an additional factor of a few (see
e.g., \cite{Murase10}).

\section{Conclusions}\label{sec:conc}
This work has examined the possibility that pulsars born with
$\sim\,$msec periods can accelerate ions to ultra-high energies at the
ultra-relativistic termination shock of the pulsar wind.  It is
motivated mainly by the observation that the Crab nebula, and other
pulsar wind nebulae, are able to accelerate pairs at a maximally
efficient rate, i.e up to the radiation reaction limit. Known pulsar
wind nebulae are however not powerful enough to confine ions up to
energies of the order of $10^{20}\,$eV. In this paper, we have
therefore constructed a phenomenological model of a much more powerful
PWN, with input rotational energy $E_{\rm rot}\,\sim\,10^{52}\,$erg
and typical spin-down time-scale $t_{\rm sd}\,\sim\,3\times 10^7\,$s;
these values correspond to a typical wind luminosity $L_{\rm
  w}\,\sim\,10^{45}\,$erg/s, a typical neutron star magnetic field
$B_\star\,\sim\,10^{13}\,$G and a typical spin-down rate $\dot
P/P\,\sim\,5\times 10^{-8}\,$/s.

Our acceleration model assumes that the wind luminosity is
substantially converted into particle random kinetic energy through
dissipation/acceleration processes around the termination shock, which
is supported by strong observational and theoretical arguments that we
recalled in Sec.~\ref{subsection:crab}. We speculate that acceleration
to the highest energies can be provided by a relativistic Fermi
mechanism at the termination shock, as envisaged in the case of known
PWNe.

We find that such PWNe are indeed able to accelerate and confine
protons up to energies of the order of the GZK cut-off, for the above
fiducial values; synchrotron energy losses are a severe limitation,
but their magnitude depends strongly on the parameters of the neutron
star.  Heavier ions could be accelerated to even larger energies, as
they are more easily confined and less affected by synchrotron energy
losses. We also find that photo-hadronic losses are not a strong
limitation of the acceleration process.

A key assumption of the present work is that ions are injected in the
pulsar wind. Such a possibility could be tested by searching for the
neutrino signal associated with photopion or photodisintegration
losses in the nebula or its
surroundings~\cite{1997PhRvL..79.2616B,2001A&A...378L..49B,2003A&A...407....1B,2003A&A...402..827A,FKMO14}. For
the fiducial parameters of the sources that we consider, namely a
luminosity $L_{\rm p}\,\sim\,10^{45}\,$erg/s and spin-down timescale
$t_{\rm sd}\,\sim\,3\times10^7\,$s, the neutrino fluxes peak at high
energies $\sim 10^{17}-10^{18}\,$eV, with a maximum diffuse flux that
lies slightly below the 5-year sensitivity of IceCube-86 and above the
3-year sensitivity of the projected ARA. It might thus become
detectable in the next decade, depending on the exact level of
contribution of these msec PWNe to the ultra-high energy cosmic ray
flux.  One should also search for neutrino signals from galactic
sources, as discussed
in~\cite{1997PhRvL..79.2616B,2001A&A...378L..49B,2003A&A...407....1B,2003A&A...402..827A,Fang14}. However,
according to the results of Ref.~\cite{2003A&A...407....1B}, the
neutrino flux of Crab-like nebulae should be at most comparable to the
atmospheric neutrino background in $10-100$~TeV range if the nuclei
take up a fraction $\eta_{\rm i}\,\sim\,0.1$ of the pulsar wind
luminosity.

Beyond the injection of ions into the pulsar wind, another central
assumption of the present work is that a fraction of pulsars are born
with a millisecond period. In order to match the cosmic-ray flux at
energies $\gtrsim 10^{19.5}\,$eV, this fraction should represent
$\sim0.03\,$\% of the present supernova rate. Although no pulsar with
a millisecond period at birth has yet been identified unambiguously,
the existence of such objects may be indirectly supported by the
recent detections of super-luminous and trans-relativistic supernovae,
which can be modelled as engine-driven supernovae, with the inner
engine being a young fast spinning pulsar injecting up to
$\sim\,10^{52}\,$ergs in the remnant on timescales as long as a few
years, see e.g.~\cite{KPO13,Metzger14, Murase14}. Actually, it has
been suggested that the Crab pulsar itself was born with a $5\,$msec
period, in order to explain the large number of radio-emitting pairs
found in the nebula~\cite{Atoyan99}. Clearly, further detailed
theoretical work and multi-wavelength follow-up of these supernovae is
needed to clarify the possible existence of such objects.

Under the above assumptions, the present work points out a new
potential acceleration mechanism of very high energy protons and this
is valuable, in our opinion, in the present context of ultra-high
energy cosmic ray physics. Indeed, on the theoretical side, there are
only a handful of sources capable of accelerating protons up to the
GZK cut-off: gamma-ray bursts~\cite{Waxman95,V95}, the most powerful
radio-galaxies~\cite{RB93}, or magnetars~\cite{Arons03}. However,
radio-galaxies with a luminosity $\gtrsim 10^{45}\,$erg/s,
sufficiently large to allow the acceleration of protons to
$10^{20}\,$eV, are too rare in the GZK sphere and the arrival
directions of the highest energy events do not match any of these, see
the discussion in Ref.~\cite{2009JCAP...11..009L}.  Gamma-ray bursts
seemingly offer the requisite conditions for the acceleration of
protons to ultra-high energies~\cite{Waxman95}, but the non-detection
of neutrinos from these sources start to constrain the amount of
energy that is injected in cosmic rays,
e.g.~\cite{2012Natur.484..351A}.  Finally, the observation of
anisotropies in the arrival directions of high energy events without
an anisotropic counterpart at lower energies does point towards the
existence of protons at GZK
energies~\cite{2009JCAP...11..009L,2013ApJ...776...88L}. Such
anisotropies have been claimed by the Pierre Auger Observatory, at a
confidence level of $99\,\%$ c.l.~\cite{2011JCAP...06..022P}, and
recently by the Telescope Array~\cite{2014arXiv1404.5890T}. In this
context, the search for sources of ultra-high energy protons, both on
the theoretical and on the experimental level, remains crucial.

\acknowledgments KK thanks Elena Amato, Pasquale Blasi, Kohta Murase
and Ke Fang for very fruitful discussions.  This work was supported by
the Programme National des Hautes Energies (CNRS).

\bibliographystyle{JHEP}
\bibliography{LKP13}

\providecommand{\href}[2]{#2}\begingroup\raggedright\begin{thebibliography}{100}

\bibitem{KO11}
K.~{Kotera} and A.~V. {Olinto}, {\it {The Astrophysics of Ultrahigh Energy
  Cosmic Rays}},  {\em ARAA} {\bf 49} (Jan., 2011)
  [\href{http://xxx.lanl.gov/abs/1101.4256}{{\tt arXiv:1101.4256}}].

\bibitem{Letessier11}
A.~{Letessier-Selvon} and T.~{Stanev}, {\it Ultrahigh energy cosmic rays},
  {\em Rev. Mod. Phys.} (2011).

\bibitem{Abbasi08}
R.~U. {Abbasi} et~al., {\it {First Observation of the Greisen-Zatsepin-Kuzmin
  Suppression}},  {\em Physical Review Letters} {\bf 100} (Mar., 2008)
  101101--+, [\href{http://xxx.lanl.gov/abs/astro-ph/}{{\tt astro-ph/}}].

\bibitem{Abraham:2008ru}
{\bf Pierre Auger} Collaboration, J.~{Abraham} et~al., {\it {Observation of the
  suppression of the flux of cosmic rays above $4\times 10^{19}$eV}},  {\em
  Phys. Rev. Lett.} {\bf 101} (2008) 061101,
  [\href{http://xxx.lanl.gov/abs/0806.4302}{{\tt arXiv:0806.4302}}].

\bibitem{2013ApJ...771...54S}
L.~{Sironi}, A.~{Spitkovsky}, and J.~{Arons}, {\it {The Maximum Energy of
  Accelerated Particles in Relativistic Collisionless Shocks}},  {\em ApJ} {\bf
  771} (July, 2013) 54, [\href{http://xxx.lanl.gov/abs/1301.5333}{{\tt
  arXiv:1301.5333}}].

\bibitem{Katz09}
B.~{Katz}, R.~{Budnik}, and E.~{Waxman}, {\it {The energy production rate and
  the generation spectrum of UHECRs}},  {\em Journal of Cosmology and
  Astro-Particle Physics} {\bf 3} (Mar., 2009) 20--+,
  [\href{http://xxx.lanl.gov/abs/0811.3759}{{\tt arXiv:0811.3759}}].

\bibitem{Bykov12}
A.~{Bykov}, N.~{Gehrels}, H.~{Krawczynski}, M.~{Lemoine}, G.~{Pelletier}, and
  M.~{Pohl}, {\it {Particle Acceleration in Relativistic Outflows}},  {\em Sp.
  Sc. Rev.} {\bf 173} (Nov., 2012) 309--339,
  [\href{http://xxx.lanl.gov/abs/1205.2208}{{\tt arXiv:1205.2208}}].

\bibitem{Waxman95}
E.~{Waxman}, {\it {Cosmological Origin for Cosmic Rays above 10 19 eV}},  {\em
  ApJl} {\bf 452} (Oct., 1995) L1+,
  [\href{http://xxx.lanl.gov/abs/astro-ph/}{{\tt astro-ph/}}].

\bibitem{V95}
M.~{Vietri}, {\it {The Acceleration of Ultra--High-Energy Cosmic Rays in
  Gamma-Ray Bursts}},  {\em ApJ} {\bf 453} (Nov., 1995) 883--+,
  [\href{http://xxx.lanl.gov/abs/astro-ph/}{{\tt astro-ph/}}].

\bibitem{Gialis04}
D.~{Gialis} and G.~{Pelletier}, {\it {Which acceleration process for ultra high
  energy cosmic rays in gamma ray bursts?}},  {\em A\&A} {\bf 425} (Oct., 2004)
  395--403, [\href{http://xxx.lanl.gov/abs/astro-ph/}{{\tt astro-ph/}}].

\bibitem{W01}
E.~{Waxman}, {\it {High-Energy Particles from {$\gamma$}-Ray Bursts}},  in {\em
  Phys Astrophys of UHECRs} (M.~{Lemoine} and G.~{Sigl}, eds.), vol.~576 of
  {\em Lecture Notes in Physics, Berlin Springer Verlag}, p.~122, 2001.

\bibitem{2010ApJ...724.1366D}
C.~D. {Dermer} and S.~{Razzaque}, {\it {Acceleration of Ultra-high-energy
  Cosmic Rays in the Colliding Shells of Blazars and Gamma-ray Bursts:
  Constraints from the Fermi Gamma-ray Space Telescope}},  {\em ApJ} {\bf 724}
  (Dec., 2010) 1366--1372, [\href{http://xxx.lanl.gov/abs/1004.4249}{{\tt
  arXiv:1004.4249}}].

\bibitem{RB93}
J.~P. {Rachen} and P.~L. {Biermann}, {\it {Extragalactic Ultra-High Energy
  Cosmic-Rays - Part One - Contribution from Hot Spots in Fr-II Radio
  Galaxies}},  {\em A\&A} {\bf 272} (May, 1993) 161--+,
  [\href{http://xxx.lanl.gov/abs/astro-ph/}{{\tt astro-ph/}}].

\bibitem{GA99}
Y.~A. {Gallant} and A.~{Achterberg}, {\it {Ultra-high-energy cosmic ray
  acceleration by relativistic blast waves}},  {\em MNRAS} {\bf 305} (May,
  1999) L6--L10, [\href{http://xxx.lanl.gov/abs/astro-ph/}{{\tt astro-ph/}}].

\bibitem{Chevalier77}
R.~A. {Chevalier}, {\it {Was SN 1054 A Type II Supernova?}},  in {\em
  Supernovae} (D.~N. {Schramm}, ed.), vol.~66 of {\em Astrophysics and Space
  Science Library}, p.~53, 1977.

\bibitem{Chevalier92}
R.~A. {Chevalier} and C.~{Fransson}, {\it {Pulsar nebulae in supernovae}},
  {\em ApJ} {\bf 395} (Aug., 1992) 540--552.

\bibitem{Gaensler05}
B.~M. {Gaensler} and P.~O. {Slane}, {\it {The Evolution and Structure of Pulsar
  Wind Nebulae}},  {\em ARAA} {\bf 44} (Sept., 2006) 17--47,
  [\href{http://xxx.lanl.gov/abs/astro-ph/}{{\tt astro-ph/}}].

\bibitem{Atoyan96}
A.~M. {Atoyan} and F.~A. {Aharonian}, {\it {On the mechanisms of gamma
  radiation in the Crab Nebula}},  {\em MNRAS} {\bf 278} (Jan., 1996) 525--541.

\bibitem{BO98}
J.~{Bednarz} and M.~{Ostrowski}, {\it {Energy Spectra of Cosmic Rays
  Accelerated at Ultrarelativistic Shock Waves}},  {\em Physical Review
  Letters} {\bf 80} (May, 1998) 3911--3914,
  [\href{http://xxx.lanl.gov/abs/astro-ph/}{{\tt astro-ph/}}].

\bibitem{Kirk00}
J.~G. {Kirk}, A.~W. {Guthmann}, Y.~A. {Gallant}, and A.~{Achterberg}, {\it
  {Particle Acceleration at Ultrarelativistic Shocks: An Eigenfunction
  Method}},  {\em ApJ} {\bf 542} (Oct., 2000) 235--242,
  [\href{http://xxx.lanl.gov/abs/astro-ph/}{{\tt astro-ph/}}].

\bibitem{Achterberg01}
A.~{Achterberg}, Y.~A. {Gallant}, J.~G. {Kirk}, and A.~W. {Guthmann}, {\it
  {Particle acceleration by ultrarelativistic shocks: theory and simulations}},
   {\em MNRAS} {\bf 328} (Dec., 2001) 393--408,
  [\href{http://xxx.lanl.gov/abs/astro-ph/}{{\tt astro-ph/}}].

\bibitem{LP03}
M.~{Lemoine} and G.~{Pelletier}, {\it {Particle Transport in Tangled Magnetic
  Fields and Fermi Acceleration at Relativistic Shocks}},  {\em ApJl} {\bf 589}
  (June, 2003) L73--L76, [\href{http://xxx.lanl.gov/abs/astro-ph/}{{\tt
  astro-ph/}}].

\bibitem{Keshet05}
U.~{Keshet} and E.~{Waxman}, {\it {Energy Spectrum of Particles Accelerated in
  Relativistic Collisionless Shocks}},  {\em Physical Review Letters} {\bf 94}
  (Mar., 2005) 111102, [\href{http://xxx.lanl.gov/abs/astro-ph/}{{\tt
  astro-ph/}}].

\bibitem{Arons03}
J.~{Arons}, {\it {Magnetars in the Metagalaxy: An Origin for Ultra-High-Energy
  Cosmic Rays in the Nearby Universe}},  {\em ApJ} {\bf 589} (June, 2003)
  871--892, [\href{http://xxx.lanl.gov/abs/astro-ph/}{{\tt astro-ph/}}].

\bibitem{Hoshino92}
M.~{Hoshino}, J.~{Arons}, Y.~A. {Gallant}, and A.~B. {Langdon}, {\it
  {Relativistic magnetosonic shock waves in synchrotron sources - Shock
  structure and nonthermal acceleration of positrons}},  {\em ApJ} {\bf 390}
  (May, 1992) 454--479.

\bibitem{Gallant94}
Y.~A. {Gallant} and J.~{Arons}, {\it {Structure of relativistic shocks in
  pulsar winds: A model of the wisps in the Crab Nebula}},  {\em ApJ} {\bf 435}
  (Nov., 1994) 230--260.

\bibitem{2004ApJ...603..669S}
A.~{Spitkovsky} and J.~{Arons}, {\it {Time Dependence in Relativistic
  Collisionless Shocks: Theory of the Variable ``Wisps'' in the Crab Nebula}},
  {\em ApJ} {\bf 603} (Mar., 2004) 669--681,
  [\href{http://xxx.lanl.gov/abs/astro-ph/0402123}{{\tt astro-ph/0402123}}].

\bibitem{Venkatesan97}
A.~{Venkatesan}, M.~C. {Miller}, and A.~V. {Olinto}, {\it {Constraints on the
  Production of Ultra--High-Energy Cosmic Rays by Isolated Neutron Stars}},
  {\em ApJ} {\bf 484} (July, 1997) 323--+,
  [\href{http://xxx.lanl.gov/abs/astro-ph/}{{\tt astro-ph/}}].

\bibitem{Blasi00}
P.~{Blasi}, R.~I. {Epstein}, and A.~V. {Olinto}, {\it {Ultra-High-Energy Cosmic
  Rays from Young Neutron Star Winds}},  {\em ApJ Letters} {\bf 533} (Apr.,
  2000) L123--L126, [\href{http://xxx.lanl.gov/abs/astro-ph/}{{\tt
  astro-ph/}}].

\bibitem{Fang12}
K.~{Fang}, K.~{Kotera}, and A.~V. {Olinto}, {\it {Newly Born Pulsars as Sources
  of Ultrahigh Energy Cosmic Rays}},  {\em The Astrophysical Journal} {\bf 750}
  (May, 2012) 118, [\href{http://xxx.lanl.gov/abs/1201.5197}{{\tt
  arXiv:1201.5197}}].

\bibitem{Fang13}
K.~{Fang}, K.~{Kotera}, and A.~V. {Olinto}, {\it {Ultrahigh energy cosmic ray
  nuclei from extragalactic pulsars and the effect of their Galactic
  counterparts}},  {\em JCAP} {\bf 3} (Mar., 2013) 10,
  [\href{http://xxx.lanl.gov/abs/1302.4482}{{\tt arXiv:1302.4482}}].

\bibitem{Kasen10}
D.~{Kasen} and L.~{Bildsten}, {\it {Supernova Light Curves Powered by Young
  Magnetars}},  {\em ApJ} {\bf 717} (July, 2010) 245--249,
  [\href{http://xxx.lanl.gov/abs/0911.0680}{{\tt arXiv:0911.0680}}].

\bibitem{Woosley10}
S.~E. {Woosley}, {\it {Bright Supernovae from Magnetar Birth}},  {\em ApJ
  Letters} {\bf 719} (Aug., 2010) L204--L207,
  [\href{http://xxx.lanl.gov/abs/0911.0698}{{\tt arXiv:0911.0698}}].

\bibitem{Dessart12}
L.~{Dessart}, D.~J. {Hillier}, R.~{Waldman}, E.~{Livne}, and S.~{Blondin}, {\it
  {Super-luminous supernovae: 56Ni power versus magnetar radiation}},  {\em
  ArXiv e-prints} (Aug., 2012) [\href{http://xxx.lanl.gov/abs/1208.1214}{{\tt
  arXiv:1208.1214}}].

\bibitem{KPO13}
K.~{Kotera}, E.~S. {Phinney}, and A.~V. {Olinto}, {\it {Signatures of pulsars
  in the light curves of newly formed supernova remnants}},  {\em MNRAS} (May,
  2013) [\href{http://xxx.lanl.gov/abs/1304.5326}{{\tt arXiv:1304.5326}}].

\bibitem{Metzger14}
B.~D. {Metzger}, I.~{Vurm}, R.~{Hasco{\"e}t}, and A.~M. {Beloborodov}, {\it
  {Ionization break-out from millisecond pulsar wind nebulae: an X-ray probe of
  the origin of superluminous supernovae}},  {\em MNRAS} {\bf 437} (Jan., 2014)
  703--720, [\href{http://xxx.lanl.gov/abs/1307.8115}{{\tt arXiv:1307.8115}}].

\bibitem{Quimby12}
R.~M. {Quimby}, {\it {Superluminous Supernovae}},  in {\em IAU Symposium},
  vol.~279 of {\em IAU Symposium}, pp.~22--28, Sept., 2012.

\bibitem{Atoyan99}
A.~M. {Atoyan}, {\it {Radio spectrum of the Crab nebula as an evidence for fast
  initial spin of its pulsar}},  {\em A\&A} {\bf 346} (June, 1999) L49--L52,
  [\href{http://xxx.lanl.gov/abs/astro-ph/9905204}{{\tt astro-ph/9905204}}].

\bibitem{Wang07}
X.-Y. {Wang}, S.~{Razzaque}, P.~{M{\'e}sz{\'a}ros}, and Z.-G. {Dai}, {\it
  {High-energy cosmic rays and neutrinos from semirelativistic hypernovae}},
  {\em Phys. Rev. D} {\bf 76} (Oct., 2007) 083009--+,
  [\href{http://xxx.lanl.gov/abs/0705.0027}{{\tt arXiv:0705.0027}}].

\bibitem{2012ApJ...746...40L}
R.-Y. {Liu} and X.-Y. {Wang}, {\it {Energy Spectrum and Chemical Composition of
  Ultrahigh Energy Cosmic Rays from Semi-relativistic Hypernovae}},  {\em ApJ}
  {\bf 746} (Feb., 2012) 40, [\href{http://xxx.lanl.gov/abs/1111.6256}{{\tt
  arXiv:1111.6256}}].

\bibitem{2011NatCo...2E.175C}
S.~{Chakraborti}, A.~{Ray}, A.~M. {Soderberg}, A.~{Loeb}, and P.~{Chandra},
  {\it {Ultra-high-energy cosmic ray acceleration in engine-driven relativistic
  supernovae}},  {\em Nature Communications} {\bf 2} (Feb., 2011)
  [\href{http://xxx.lanl.gov/abs/1012.0850}{{\tt arXiv:1012.0850}}].

\bibitem{Rees74}
M.~J. {Rees} and J.~E. {Gunn}, {\it {The origin of the magnetic field and
  relativistic particles in the Crab Nebula}},  {\em MNRAS} {\bf 167} (Apr.,
  1974) 1--12.

\bibitem{Bednarek02}
W.~{Bednarek} and R.~J. {Protheroe}, {\it {Contribution of nuclei accelerated
  by gamma-ray pulsars to cosmic rays in the Galaxy}},  {\em Astroparticle
  Physics} {\bf 16} (Feb., 2002) 397--409,
  [\href{http://xxx.lanl.gov/abs/astro-ph/}{{\tt astro-ph/}}].

\bibitem{Kennel84}
C.~F. {Kennel} and F.~V. {Coroniti}, {\it {Confinement of the Crab pulsar's
  wind by its supernova remnant}},  {\em ApJ} {\bf 283} (Aug., 1984) 694--709.

\bibitem{Kennel84b}
C.~F. {Kennel} and F.~V. {Coroniti}, {\it {Magnetohydrodynamic model of Crab
  nebula radiation}},  {\em ApJ} {\bf 283} (Aug., 1984) 710--730.

\bibitem{Komissarov04}
S.~{Komissarov} and Y.~{Lyubarsky}, {\it {MHD Simulations of Crab's Jet and
  Torus}},  {\em Ap \& SS} {\bf 293} (Sept., 2004) 107--113.

\bibitem{Porth13}
O.~{Porth}, S.~S. {Komissarov}, and R.~{Keppens}, {\it {Solution to the sigma
  problem of pulsar wind nebulae}},  {\em MNRAS} {\bf 431} (Apr., 2013)
  L48--L52, [\href{http://xxx.lanl.gov/abs/1212.1382}{{\tt arXiv:1212.1382}}].

\bibitem{Metzger11}
B.~D. {Metzger}, D.~{Giannios}, and S.~{Horiuchi}, {\it {Heavy nuclei
  synthesized in gamma-ray burst outflows as the source of ultrahigh energy
  cosmic rays}},  {\em MNRAS} {\bf 415} (Aug., 2011) 2495--2504,
  [\href{http://xxx.lanl.gov/abs/1101.4019}{{\tt arXiv:1101.4019}}].

\bibitem{DelZanna04}
L.~{Del Zanna}, E.~{Amato}, and N.~{Bucciantini}, {\it {Axially symmetric
  relativistic MHD simulations of Pulsar Wind Nebulae in Supernova Remnants. On
  the origin of torus and jet-like features}},  {\em A\&A} {\bf 421} (July,
  2004) 1063--1073, [\href{http://xxx.lanl.gov/abs/astro-ph/}{{\tt
  astro-ph/}}].

\bibitem{Gelfand09}
J.~D. {Gelfand}, P.~O. {Slane}, and W.~{Zhang}, {\it {A Dynamical Model for the
  Evolution of a Pulsar Wind Nebula Inside a Nonradiative Supernova Remnant}},
  {\em ApJ} {\bf 703} (Oct., 2009) 2051--2067,
  [\href{http://xxx.lanl.gov/abs/0904.4053}{{\tt arXiv:0904.4053}}].

\bibitem{Bucciantini11}
N.~{Bucciantini}, J.~{Arons}, and E.~{Amato}, {\it {Modelling spectral
  evolution of pulsar wind nebulae inside supernova remnants}},  {\em MNRAS}
  {\bf 410} (Jan., 2011) 381--398.

\bibitem{2014AN....335..318G}
J.~D. {Gelfand}, P.~O. {Slane}, and T.~{Temim}, {\it {The properties of the
  progenitor, neutron star, and pulsar wind in the supernova remnant Kes 75}},
  {\em Astronomische Nachrichten} {\bf 335} (Mar., 2014) 318--323.

\bibitem{2014MNRAS.443..138M}
J.~{Martin}, D.~F. {Torres}, A.~{Cillis}, and E.~{de O{\~n}a Wilhelmi}, {\it
  {Is there room for highly magnetized pulsar wind nebulae among those
  non-detected at TeV?}},  {\em MNRAS} {\bf 443} (Sept., 2014) 138--145,
  [\href{http://xxx.lanl.gov/abs/1406.1344}{{\tt arXiv:1406.1344}}].

\bibitem{Komissarov13}
S.~S. {Komissarov}, {\it {Magnetic dissipation in the Crab nebula}},  {\em
  MNRAS} {\bf 428} (Jan., 2013) 2459--2466,
  [\href{http://xxx.lanl.gov/abs/1207.3192}{{\tt arXiv:1207.3192}}].

\bibitem{Coroniti90}
F.~V. {Coroniti}, {\it {Magnetically striped relativistic magnetohydrodynamic
  winds - The Crab Nebula revisited}},  {\em ApJ} {\bf 349} (Feb., 1990)
  538--545.

\bibitem{Chiueh98}
T.~{Chiueh}, Z.-Y. {Li}, and M.~C. {Begelman}, {\it {A Critical Analysis of
  Ideal Magnetohydrodynamic Models for Crab-like Pulsar Winds}},  {\em ApJ}
  {\bf 505} (Oct., 1998) 835--843.

\bibitem{Contopoulos02}
I.~{Contopoulos} and D.~{Kazanas}, {\it {Toward Resolving the Crab
  {$\sigma$}-Problem: A Linear Accelerator?}},  {\em ApJ} {\bf 566} (Feb.,
  2002) 336--342, [\href{http://xxx.lanl.gov/abs/astro-ph/}{{\tt astro-ph/}}].

\bibitem{Kirk09}
J.~G. {Kirk}, Y.~{Lyubarsky}, and J.~{Petri}, {\it {The Theory of Pulsar Winds
  and Nebulae}},  in {\em Astrophysics and Space Science Library} (W.~{Becker},
  ed.), vol.~357 of {\em Astrophysics and Space Science Library}, p.~421, 2009.
\newblock \href{http://xxx.lanl.gov/abs/astro-ph/}{{\tt astro-ph/}}.

\bibitem{Lyubarsky10}
Y.~{Lyubarsky}, {\it {A New Mechanism for Dissipation of Alternating Fields in
  Poynting-dominated Outflows}},  {\em ApJL} {\bf 725} (Dec., 2010) L234--L238,
  [\href{http://xxx.lanl.gov/abs/1012.1411}{{\tt arXiv:1012.1411}}].

\bibitem{LP10}
M.~{Lemoine} and G.~{Pelletier}, {\it {On electromagnetic instabilities at
  ultra-relativistic shock waves}},  {\em MNRAS} {\bf 402} (Feb., 2010)
  321--334, [\href{http://xxx.lanl.gov/abs/0904.2657}{{\tt arXiv:0904.2657}}].

\bibitem{LP11}
M.~{Lemoine} and G.~{Pelletier}, {\it {Dispersion and thermal effects on
  electromagnetic instabilities in the precursor of relativistic shocks}},
  {\em MNRAS} {\bf 417} (Oct., 2011) 1148--1161,
  [\href{http://xxx.lanl.gov/abs/1102.1308}{{\tt arXiv:1102.1308}}].

\bibitem{LPR06}
M.~{Lemoine}, G.~{Pelletier}, and B.~{Revenu}, {\it {On the Efficiency of Fermi
  Acceleration at Relativistic Shocks}},  {\em ApJl} {\bf 645} (July, 2006)
  L129--L132, [\href{http://xxx.lanl.gov/abs/astro-ph/}{{\tt astro-ph/}}].

\bibitem{LPGP13}
M.~{Lemoine}, G.~{Pelletier}, L.~{Gremillet}, and I.~{Plotnikov}, {\it
  {Current-driven filamentation upstream of magnetized relativistic
  collisionless shocks}},  {\em Mont. Not. Roy. Astron. Soc.} (Mar., 2014)
  [\href{http://xxx.lanl.gov/abs/1401.7166}{{\tt arXiv:1401.7166}}].

\bibitem{Pelletier09}
G.~{Pelletier}, M.~{Lemoine}, and A.~{Marcowith}, {\it {On Fermi acceleration
  and magnetohydrodynamic instabilities at ultra-relativistic magnetized shock
  waves}},  {\em MNRAS} {\bf 393} (Feb., 2009) 587--597,
  [\href{http://xxx.lanl.gov/abs/0807.3459}{{\tt arXiv:0807.3459}}].

\bibitem{Michel82}
F.~C. {Michel}, {\it {Theory of pulsar magnetospheres}},  {\em Reviews of
  Modern Physics} {\bf 54} (Jan., 1982) 1--66.

\bibitem{Michel94}
F.~C. {Michel}, {\it {Magnetic structure of pulsar winds}},  {\em ApJ} {\bf
  431} (Aug., 1994) 397--401.

\bibitem{Kirk03}
J.~G. {Kirk} and O.~{Skj{\ae}raasen}, {\it {Dissipation in
  Poynting-Flux-dominated Flows: The {$\sigma$}-Problem of the Crab Pulsar
  Wind}},  {\em ApJ} {\bf 591} (July, 2003) 366--379,
  [\href{http://xxx.lanl.gov/abs/astro-ph/}{{\tt astro-ph/}}].

\bibitem{BS73}
A.~{Barnes} and J.~D. {Scargle}, {\it {Collisionless Damping of Hydromagnetic
  Waves in Relativistic Plasma. Weak Landau Damping}},  {\em ApJ} {\bf 184}
  (Aug., 1973) 251--270.

\bibitem{Lyubarsky03}
Y.~E. {Lyubarsky}, {\it {The termination shock in a striped pulsar wind}},
  {\em MNRAS} {\bf 345} (Oct., 2003) 153--160,
  [\href{http://xxx.lanl.gov/abs/astro-ph/}{{\tt astro-ph/}}].

\bibitem{Petri07}
J.~{P{\'e}tri} and Y.~{Lyubarsky}, {\it {Magnetic reconnection at the
  termination shock in a striped pulsar wind}},  {\em A\&A} {\bf 473} (Oct.,
  2007) 683--700.

\bibitem{Sironi11}
L.~{Sironi} and A.~{Spitkovsky}, {\it {Acceleration of Particles at the
  Termination Shock of a Relativistic Striped Wind}},  {\em ApJ} {\bf 741}
  (Nov., 2011) 39, [\href{http://xxx.lanl.gov/abs/1107.0977}{{\tt
  arXiv:1107.0977}}].

\bibitem{Camus09}
N.~F. {Camus}, S.~S. {Komissarov}, N.~{Bucciantini}, and P.~A. {Hughes}, {\it
  {Observations of `wisps' in magnetohydrodynamic simulations of the Crab
  Nebula}},  {\em MNRAS} {\bf 400} (Dec., 2009) 1241--1246,
  [\href{http://xxx.lanl.gov/abs/0907.3647}{{\tt arXiv:0907.3647}}].

\bibitem{Hibschman01}
J.~A. {Hibschman} and J.~{Arons}, {\it {Pair Production Multiplicities in
  Rotation-powered Pulsars}},  {\em ApJ} {\bf 560} (Oct., 2001) 871--884,
  [\href{http://xxx.lanl.gov/abs/astro-ph/0107209}{{\tt astro-ph/0107209}}].

\bibitem{Timokhin10}
A.~N. {Timokhin}, {\it {Time-dependent pair cascades in magnetospheres of
  neutron stars - I. Dynamics of the polar cap cascade with no particle supply
  from the neutron star surface}},  {\em MNRAS} {\bf 408} (Nov., 2010)
  2092--2114, [\href{http://xxx.lanl.gov/abs/1006.2384}{{\tt
  arXiv:1006.2384}}].

\bibitem{Reynolds84}
S.~P. {Reynolds} and R.~A. {Chevalier}, {\it {Evolution of pulsar-driven
  supernova remnants}},  {\em ApJ} {\bf 278} (Mar., 1984) 630--648.

\bibitem{2014JHEAp...1...31T}
D.~F. {Torres}, A.~{Cillis}, J.~{Mart{\'{\i}}n}, and E.~{de O{\~n}a Wilhelmi},
  {\it {Time-dependent modeling of TeV-detected, young pulsar wind nebulae}},
  {\em Journal of High Energy Astrophysics} {\bf 1} (May, 2014) 31--62,
  [\href{http://xxx.lanl.gov/abs/1402.5485}{{\tt arXiv:1402.5485}}].

\bibitem{Goldreich69}
P.~{Goldreich} and W.~H. {Julian}, {\it {Pulsar Electrodynamics}},  {\em ApJ}
  {\bf 157} (Aug., 1969) 869--+.

\bibitem{2014PhRvD..90l2006A}
A.~{Aab} and et~al. (Pierre Auger~Collaboration), {\it {Depth of maximum of
  air-shower profiles at the Pierre Auger Observatory. II. Composition
  implications}},  {\em Phys. Rev. D} {\bf 90} (Dec., 2014) 122006.

\bibitem{2015APh....64...49A}
R.~U. {Abbasi} and et~al. (Telescope Array~Collaboration), {\it {Study of
  Ultra-High Energy Cosmic Ray composition using Telescope Array's Middle Drum
  detector and surface array in hybrid mode}},  {\em Astroparticle Physics}
  {\bf 64} (Apr., 2015) 49--62, [\href{http://xxx.lanl.gov/abs/1408.1726}{{\tt
  arXiv:1408.1726}}].

\bibitem{2015arXiv150307540A}
R.~{Abbasi}, J.~{Bellido}, J.~{Belz}, V.~{de Souza}, W.~{Hanlon}, D.~{Ikeda},
  J.~P. {Lundquist}, P.~{Sokolsky}, T.~{Stroman}, Y.~{Tameda}, Y.~{Tsunesada},
  M.~{Unger}, {A.~Yushkov for the Pierre Auger Collaboration}, and {the
  Telescope Array Collaboration}, {\it {Report of the Working Group on the
  Composition of Ultra High Energy Cosmic Rays}},  {\em ArXiv e-prints} (Mar.,
  2015) [\href{http://xxx.lanl.gov/abs/1503.0754}{{\tt arXiv:1503.0754}}].

\bibitem{Niemiec06}
J.~{Niemiec}, M.~{Ostrowski}, and M.~{Pohl}, {\it {Cosmic-Ray Acceleration at
  Ultrarelativistic Shock Waves: Effects of Downstream Short-Wave Turbulence}},
   {\em ApJ} {\bf 650} (Oct., 2006) 1020--1027,
  [\href{http://xxx.lanl.gov/abs/astro-ph/}{{\tt astro-ph/}}].

\bibitem{1976PhFl...19.1130B}
R.~D. {Blandford} and C.~F. {McKee}, {\it {Fluid dynamics of relativistic blast
  waves}},  {\em Physics of Fluids} {\bf 19} (Aug., 1976) 1130--1138.

\bibitem{SS99}
F.~W. {Stecker} and M.~H. {Salamon}, {\it {Photodisintegration of
  Ultra-High-Energy Cosmic Rays: A New Determination}},  {\em ApJ} {\bf 512}
  (Feb., 1999) 521--526, [\href{http://xxx.lanl.gov/abs/astro-ph/}{{\tt
  astro-ph/}}].

\bibitem{1998PhRvD..58l3005R}
J.~P. {Rachen} and P.~{M{\'e}sz{\'a}ros}, {\it {Photohadronic neutrinos from
  transients in astrophysical sources}},  {\em Phys. Rev. D} {\bf 58} (Dec.,
  1998) 123005, [\href{http://xxx.lanl.gov/abs/astro-ph/9802280}{{\tt
  astro-ph/9802280}}].

\bibitem{Abbasi:2011zx}
R.~Abbasi et~al., {\it {The IceCube Neutrino Observatory II: All Sky Searches:
  Atmospheric, Diffuse and EHE}},
  \href{http://xxx.lanl.gov/abs/1111.2736}{{\tt arXiv:1111.2736}}.

\bibitem{Adams:2012tt}
{\bf JEM-EUSO} Collaboration, J.~Adams et~al., {\it {The JEM-EUSO Mission:
  Status and Prospects in 2011}},
  \href{http://xxx.lanl.gov/abs/1204.5065}{{\tt arXiv:1204.5065}}.

\bibitem{Allison12}
P.~Allison, J.~Auffenberg, R.~Bard, J.~Beatty, D.~Besson, et~al., {\it {Design
  and Initial Performance of the Askaryan Radio Array Prototype EeV Neutrino
  Detector at the South Pole}},  {\em Astropart.Phys.} {\bf 35} (2012)
  457--477, [\href{http://xxx.lanl.gov/abs/1105.2854}{{\tt arXiv:1105.2854}}].

\bibitem{ThePierreAuger:2013eja}
{\bf Pierre Auger Collaboration} Collaboration, A.~Aab et~al., {\it {The Pierre
  Auger Observatory: Contributions to the 33rd International Cosmic Ray
  Conference (ICRC 2013)}},  \href{http://xxx.lanl.gov/abs/1307.5059}{{\tt
  arXiv:1307.5059}}.

\bibitem{Murase10}
K.~{Murase} and J.~F. {Beacom}, {\it {Neutrino background flux from sources of
  ultrahigh-energy cosmic-ray nuclei}},  {\em Phys. Rev. D} {\bf 81} (June,
  2010) 123001--+, [\href{http://xxx.lanl.gov/abs/1003.4959}{{\tt
  arXiv:1003.4959}}].

\bibitem{1997PhRvL..79.2616B}
W.~{Bednarek} and R.~J. {Protheroe}, {\it {Gamma Rays and Neutrinos from the
  Crab Nebula Produced by Pulsar Accelerated Nuclei}},  {\em Phys. Rev. Lett.}
  {\bf 79} (Oct., 1997) 2616--2619,
  [\href{http://xxx.lanl.gov/abs/astro-ph/9704186}{{\tt astro-ph/9704186}}].

\bibitem{2001A&A...378L..49B}
W.~{Bednarek}, {\it {Extragalactic neutrino background from very young pulsars
  surrounded by supernova envelopes}},  {\em Astron. Astrophys.} {\bf 378}
  (Oct., 2001) L49--L52, [\href{http://xxx.lanl.gov/abs/astro-ph/0109225}{{\tt
  astro-ph/0109225}}].

\bibitem{2003A&A...407....1B}
W.~{Bednarek}, {\it {Neutrinos from the pulsar wind nebulae}},  {\em Astron.
  Astrophys.} {\bf 407} (Aug., 2003) 1--6,
  [\href{http://xxx.lanl.gov/abs/astro-ph/0305430}{{\tt astro-ph/0305430}}].

\bibitem{2003A&A...402..827A}
E.~{Amato}, D.~{Guetta}, and P.~{Blasi}, {\it {Signatures of high energy
  protons in pulsar winds}},  {\em Astron. Astrophys.} {\bf 402} (May, 2003)
  827--836, [\href{http://xxx.lanl.gov/abs/astro-ph/0302121}{{\tt
  astro-ph/0302121}}].

\bibitem{FKMO14}
K.~{Fang}, K.~{Kotera}, K.~{Murase}, and A.~V. {Olinto}, {\it {Testing the
  newborn pulsar origin of ultrahigh energy cosmic rays with EeV neutrinos}},
  {\em Phys. Rev. D} {\bf 90} (Nov., 2014) 103005,
  [\href{http://xxx.lanl.gov/abs/1311.2044}{{\tt arXiv:1311.2044}}].

\bibitem{Fang14}
K.~{Fang}, {\it {High-Energy Neutrino Signatures of Newborn Pulsars In the
  Local Universe}},  {\em ArXiv e-prints} (Nov., 2014)
  [\href{http://xxx.lanl.gov/abs/1411.2174}{{\tt arXiv:1411.2174}}].

\bibitem{Murase14}
K.~{Murase}, K.~{Kashiyama}, K.~{Kiuchi}, and I.~{Bartos}, {\it {Gamma-Ray and
  Hard X-Ray Emission from Pulsar-Aided Supernovae as a Probe of Particle
  Acceleration in Embryonic Pulsar Wind Nebulae}},  {\em ArXiv e-prints:
  1411.0619} (Nov., 2014) [\href{http://xxx.lanl.gov/abs/1411.0619}{{\tt
  arXiv:1411.0619}}].

\bibitem{2009JCAP...11..009L}
M.~{Lemoine} and E.~{Waxman}, {\it {Anisotropy vs chemical composition at
  ultra-high energies}},  {\em J. Cosm. Astropart. Phys.} {\bf 11} (Nov., 2009)
  9, [\href{http://xxx.lanl.gov/abs/0907.1354}{{\tt arXiv:0907.1354}}].

\bibitem{2012Natur.484..351A}
R.~{Abbasi}, Y.~{Abdou}, T.~{Abu-Zayyad}, M.~{Ackermann}, J.~{Adams}, J.~A.
  {Aguilar}, M.~{Ahlers}, D.~{Altmann}, K.~{Andeen}, J.~{Auffenberg}, and
  et~al., {\it {An absence of neutrinos associated with cosmic-ray acceleration
  in {$\gamma$}-ray bursts}},  {\em Nature} {\bf 484} (Apr., 2012) 351--354,
  [\href{http://xxx.lanl.gov/abs/1204.4219}{{\tt arXiv:1204.4219}}].

\bibitem{2013ApJ...776...88L}
R.-Y. {Liu}, A.~M. {Taylor}, M.~{Lemoine}, X.-Y. {Wang}, and E.~{Waxman}, {\it
  {Constraints on the Source of Ultra-high-energy Cosmic Rays Using Anisotropy
  versus Chemical Composition}},  {\em Astrophys. J.} {\bf 776} (Oct., 2013)
  88, [\href{http://xxx.lanl.gov/abs/1308.5699}{{\tt arXiv:1308.5699}}].

\bibitem{2011JCAP...06..022P}
{Pierre Auger Collaboration}, P.~{Abreu}, M.~{Aglietta}, E.~J. {Ahn}, I.~F.~M.
  {Albuquerque}, D.~{Allard}, I.~{Allekotte}, J.~{Allen}, P.~{Allison},
  J.~{Alvarez Castillo}, and et~al., {\it {Anisotropy and chemical composition
  of ultra-high energy cosmic rays using arrival directions measured by the
  Pierre Auger Observatory}},  {\em J. Cosm. Astropart. Phys.} {\bf 6} (June,
  2011) 22, [\href{http://xxx.lanl.gov/abs/1106.3048}{{\tt arXiv:1106.3048}}].

\bibitem{2014arXiv1404.5890T}
{The Telescope Array Collaboration}, R.~U. {Abbasi}, and et~al., {\it
  {Indications of Intermediate-Scale Anisotropy of Cosmic Rays with Energy
  Greater Than 57 EeV in the Northern Sky Measured with the Surface Detector of
  the Telescope Array Experiment}},  {\em ArXiv e-prints} (Apr., 2014)
  [\href{http://xxx.lanl.gov/abs/1404.5890}{{\tt arXiv:1404.5890}}].

\end{thebibliography}\endgroup

\end{document}